# Spin transport in Si-based spin metal-oxide-semiconductor field-effect transistors: Spin drift effect in the inversion channel and spin relaxation in the $n^+$-Si source/drain regions

Shoichi Sato[1,2], Masaaki Tanaka[1,2], and Ryosho Nakane[1]

[1]*Department of Electrical Engineering and Information Systems, The University of Tokyo, 7-3-1 Hongo, Bunkyo-ku, Tokyo 113-8656, Japan*

[2]*Center for Spintronics Research Network (CSRN), The University of Tokyo, 7-3-1 Hongo, Bunkyo-ku, Tokyo 113-8656, Japan*

## Abstract

We have experimentally and theoretically investigated the electron spin transport and spin distribution at room temperature in a Si two-dimensional (2D) inversion channel of back-gate-type spin metal-oxide-semiconductor field-effect transistors (spin MOSFETs). The magnetoresistance ratio of the spin MOSFET with a channel length of 0.4 μm was increased by a factor of 6 from that in our previous paper [Phys. Rev. B **99**, 165301 (2019)] by lowering the parasitic resistances at the source/drain junctions with highly-phosphorus-doped $n^+$-Si regions and by increasing the lateral electric field in the channel along the electron transport, called "spin drift". Clear Hanle signals with some oscillation peaks were observed for the spin MOSFET with a channel length of 10 μm under the lateral electric field, indicating that the effective spin diffusion length is dramatically enhanced by the spin drift. By taking into account the $n^+$-Si regions and the spin drift in the channel, one-dimensional analytic functions were derived for analyzing the effect of the spin drift on the spin transport through the channel and these functions were found to explain almost all the experimental results.

From the calculated spin current and spin distribution, it was revealed that almost all the spins are unflipped during the spin-drift-assisted transport through the 0.4-μm-long inversion channel, but the most part of the *injected spins* from the source electrode are relaxed in the $n^+$-Si regions of both the source and drain junctions. This means that the spin drift is useful and precise design of the device structure is essential to obtain a higher magnetoresistance ratio. Furthermore, we showed that the effective spin resistances that are introduced in this study are very helpful to understand how to improve the magnetoresistance ratio of spin MOSFETs for practical use by optimizing the source/drain junctions and channel structure. The most remarkable finding is that the design guideline for spin MOSFETs utilizing electron spin transport is different from that for the ordinary MOSFETs utilizing electron charge transport.

## I. Introduction

In the past decade, Si-based spin metal-oxide-semiconductor field-effect transistors (spin MOSFETs) [1-8] have been extensively studied since they possess the possibility of becoming key devices in next-generation electronics due to their spin-functional nonvolatile/reconfigurable characteristics. Spin MOSFETs have basically the same device structure as the ordinary MOSFETs, but the source and drain (S/D) electrodes are replaced with a ferromagnetic material or a ferromagnetic tunnel junction. The most unique and useful feature is that their transconductance can be changed by the magnetization configuration of the ferromagnetic S/D electrodes, which is realized by the injection of spin-polarized electrons (spin injection) at the S junction, electron conduction via the 2-dimensional (2D) channel with no spin-flip, and the detection of spin-polarized electrons (spin detection) at the D junction. The change in the transconductance is evaluated by a magnetoresistance (MR) ratio (or a magneto-current ratio) that is defined by the normalized resistance (or current) change between the parallel and antiparallel magnetization configurations in the S/D electrodes. Although some previous experimental studies showed the basic operation with a clear transistor action and a spin-valve signal, the MR ratio was lower than 1% [5-8], which is too small for practical applications. To improve the MR ratio, it is necessary to clarify the detailed spin transport physics in Si-based spin MOSFETs and to design the device structure appropriately.

In our recent paper [8], we experimentally demonstrated the spin MOSFET operation in back-gate-type Schottky-barrier spin MOSFETs with Fe-based ferromagnetic S/D electrodes and a uniformly phosphorus-doped *n*-type Si channel (the doping concentration was $10^{17}$ $cm^{-3}$), and quantitatively analyzed the electron spin transport through the 2D channel. It was found that the probability ($1/\tau_S$) of spin flip scattering is 1/14000 of the probability ($1/\tau$) of electron momentum scattering and this ratio remains unchanged by the gate electric field vertically applied to the electron transport channel. The remaining issues are i) the MR ratio of only 0.003% whih is smaller than another group's value (~0.03%) [7], and ii) the relatively poor transistor characteristics compared with those of an ordinary MOSFET with the same channel width and length. From a simple consideration, one of the origins of these poor characteristics are the large parasitic resistances arising from the Schottky barrier formed under the S/D electrodes. To improve these two device characteristics at the same time, it is required to form highly-phosphorus-doped $n^+$-Si regions under the S/D electrodes as ordinary MOSFETs. When this method is employed, since electron spins are significantly relaxed in the $n^+$-Si regions, it is necessary to examine whether the channel structure consisting of $n^+$-Si regions and a 2D channel is really effective or not, by performing systematic experiments and by making a suitable theoretical model which takes into account the detailed channel structure.

On the other hand, enhancing the effective spin diffusion length $\lambda_S$ in a 2D channel is another method to improve the MR ratio. Recent studies showed that the electron spin lifetime $\tau_S$ in a 2D electron accumulation/inversion channel is ~1 ns at room temperature [6,8,9], which is significantly smaller than that in bulk Si materials (10–100 ns) [10]. The reduction in $\tau_S$ leads to a small intrinsic spin diffusion length $\lambda_S = \sqrt{D_e \tau_S}$ of ~1 μm, which means that a channel length less than a few-tens nm is required to achieve electron transport with almost no spin-flips in the channel. In principle, enhancing $\lambda_S$ can be achieved by reducing the probability of spin-flip scattering during the electron transport through a channel. Since the $\tau_S$ value is almost unchanged by the gate electric field vertically applied to the 2D channel and the probability of spin-flip scattering is governed by the probability of electron momentum scattering [8,11-13], one feasible method for enhancing $\lambda_S$ is to increase the drift velocity $v_d$ of electrons by increasing the electric field along the electron transport. This is called "spin drift" [14]. So far, there have been some experimental studies on the spin drift in four-terminal device structures [15-17], in which the theoretical model for the signal analysis assumed a uniform electric field in the channel along the electron transport. However, this assumption is not realistic because electric fields are present even in the nonlocal regions (where no bias is applied) outside the S and D electrodes. Our recent report on spin MOSFETs [18] showed that the spin drift obtained in a wide range of bias conditions can be consistently explained by our theoretical model with an assumption that a uniform electric field is present only in the local regions between the S and D electrodes. Thus, it is very important to take into account the realistic channel structure in detailed analyses of the spin transport.

In this study, we prepare Si-based back-gate-type spin MOSFETs with highly-phosphorus-doped $n^+$-Si regions beneath the ferromagnetic S/D electrodes (see Fig. 1), and systematically measure spin-valve and Hanle precession signals at room temperature with various bias currents and gate voltages. To theoritically investigate the spin transport, we construct analytical formulas that precisely take into account the channel structure consisting of a Si inversion channel sandwiched by two $n^+$-Si regions as well as the spin drift effect in the inversion channel. The spin transport from the S to D electrodes and appearance of magnetoresistance are found to be strongly influenced by the relationship between the gate-electric-field-, bias-, and direction-dependent spin resistance in the inversion channel and a constant spin resistance in the $n^+$-Si region. Our results help deep understanding of the spin transport through a practical channel structure, which is essential for designing high-performance spin MOSFETs with both high MR ratios and good transistor characteristics.

Section II describes the preparation of spin MOSFETs which have Fe(4 nm) / Mg(2 nm) / MgO (1 nm) / $n^+$-Si(5 nm) junctions at the S/D electrodes and an 8-nm-thick $p$-type Si channel with channel lengths $L_{ch}$ = 0.4 and 10 μm. Section III shows experimental results, such as transistor characteristics, spin-valve signals, and Hanle precession signals at room temperature. In Section IV, we present our original model that takes into account both the $n^+$-Si regions at the S/D junctions and the lateral electric field in the Si inversion channel, and then analyze the experimental signals. Section V shows the spin distribution in our spin MOSFETs and discusses how the spin drift and the $n^+$-Si region determine the MR ratio. In Section VI, we describe concluding remarks and address future issues.

## II. Device preparation

Figures 1(a) and (b) show the side and top views of our spin MOSFET structure, respectively, which has ferromagnetic S/D electrodes and two reference electrodes (R1 and R2) located outside of the S and D electrodes, respectively. Since the device fabrication process is almost the same as that in our recent study [8] except the substrate preparation at the early stage of the process, the difference is mainly described here. First, a substrate was cut from a (001)-oriented silicon-on-insulator (SOI) wafer whose structure is (from top to bottom) a 100-nm-thick *p*-type Si layer with an accepter doping concentration of ~$10^{15}$ $cm^{-3}$, a buried silicon-dioxide (BOX) layer with a thickness $t_{BOX}$ = 200 nm, and a *p*-type Si substrate. The substrate was thermally oxidized at 1050°C with dry oxygen gas to form a 220-nm-thick surface $SiO_2$ layer and a 8-nm-thick Si channel layer. The surface $SiO_2$ layer was thinned to 120 nm by buffered HF (BHF). Using electron beam (EB) lithography, S, D, R1, and R2 contact areas were defined by the EB resist, and then the surface $SiO_2$ was etched with BHF so that the $SiO_2$ thickness of these areas becomes ~20 nm. After removing the EB resist, the substrate was cleaned by $H_2SO_4$ and $H_2O_2$ mixture (SPM), followed by flowing de-ionized (DI) water, and then the Si surface of the contact areas was opened by dipping the whole substrate into BHF for a few seconds. Next, these contact areas were doped with phosphorous donor atoms using a spin-coated phosphorous glass ($P_2O_5$) and thermal diffusion at 970°C for 5 min to form $n^+$-Si regions (shown by yellow regions in Fig. 1(a)). Note that the $n^+$-Si regions reach the bottom of the Si layer to eliminate the Schottky-barrier resistance beneath the S/D electrodes. The phosphorous donor concentration $N_D$ in the $n^+$-Si regions was estimated to be ~$1\times10^{20}$ cm$^{-3}$ by a Hall measurement. After removing $P_2O_5$ by 1% HF, the substrate was cleaned again by SPM, followed by flowing DI water. During the second SPM cleaning, the thickness of the $n^+$-Si region was decreased to $t_n$ = 5 nm. The thicknesses of the $p$-Si channel ($t_{Si}$ = 8 nm) and the $n^+$-Si regions ($t_n$ = 5 nm) were confirmed by a cross-sectional transmission electron microscopy (TEM) observation as shown in Fig. 1(c). Then, the substrate was installed into an ultra-high vacuum system, and (from top to bottom) an Al(15 nm)/Mg(1 nm)/Fe(4 nm)/Mg(2 nm)/MgO(1 nm) layered structure was successively deposited at room temperature by molecular beam epitaxy (MBE) and EB evaporation. After the deposition, a back-gate-type MOSFET structure was fabricated by the same procedure described in ref. [8]. The channel length $L_{ch}$ along the $y$ axis ($[1\bar{1}0]$ axis of Si) are 0.4 or 10 μm, the channel width $W_{ch}$ along the $x$ axis ([110] axis of Si) is 180 μm, the distance between the S and R1 (D and R2) electrodes along the $y$ axis is ~100 μm, the lengths of the S electrode $L_S$ and the D electrode $L_D$ along the $y$ axis are 0.7 and 2.0 μm, respectively, and the lengths of the R1 and R2 electrodes along the $y$ axis are 40 μm. Hereafter, the ferromagnetic Al/Mg/Fe/Mg/MgO/$n^+$-Si tunnel junctions at the S and D are referred to as "the S junction" and "the D junction", respectively.

Figure 1(d) shows our measurement setup for the two-terminal (2T) spin transport signals, where the voltages ($V_{DS}$ and $V_D$) between the two electrodes are measured while a constant current $I_{DS}$ is driven from the D to S electrodes (electrons flow from the S to D electrodes) through the Si channel, a constant gate-source voltage $V_{GS}$ is applied between the back-gate electrode and S electrode, and a magnetic field $H$ is applied. The direction of $H$ is parallel to the $x$ direction ($H = H_{//}$) for spin-valve signal measurements, while it is parallel to the $z$ direction ($H = H_{\perp}$) for Hanle precession signal measurements. $V_{DS}$ is the voltage between the D and S electrodes, which is the total voltage drop through the Si channel and the D/S junctions, and $V_D$ is the voltage between D and R2, which is basically the junction voltage drop of the MgO tunnel barrier at the D junction. All the measurements were performed at 295 K.

## III. Experimental results

### A. Spin MOSFET operations in the device with $L_{ch}$ = 0.4 μm

Figures 2(a) and (b) show $I_{DS}$ - $V_{DS}$ and $I_{DS}$ - $V_{GS}$ characteristics measured for a device with $L_{ch}$ = 0.4 μm, respectively, which confirm clear transistor operations with a high on-off ratio ~$10^6$. Owing to the $n^+$-Si regions at the S/D junctions, the parasitic resistance was reduced compared with our previous result of the spin MOSFET without $n^+$-Si regions [8]; the linearity in the $I_{DS}$ - $V_{DS}$ characteristics is improved and the $I_{DS}$ is larger by ~2 times than the Schottky spin MOSFET with the same channel length. Figures 2(c) and (d) show the voltage change $\Delta V_{DS}$ as a function of $H_{//}$ for the same device, which were measured (c) with various $I_{DS}$ (= 2 – 10 mA) and a constant $V_{GS}$ = 40 V, and (d) with various $V_{GS}$ (= 40 – 60 V) and a constant $I_{DS}$ = 10 mA. The green solid and dashed curves in Fig. 2(c) are the major and the minor loop characteristics measured with $I_{DS}$ = 5 mA and $V_{GS}$ = 40 V, which verify that these signals are originated from the spin transport from the S to D electrodes. This means that the spin-valve effect was clearly observed, which is caused by the parallel and antiparallel magnetization configurations of the S and D electrodes. Thus, the basic spin MOSFETs operation was demonstrated at room temperature. Here, we define the amplitude of the spin-valve signal $\Delta V^{2T}$ as shown in Fig. 2(c). It is found that $\Delta V^{2T}$ increases with increasing $I_{DS}$ in Fig. 2(c), whereas it increases with decreasing $V_{GS}$ in Fig. 2(d). These $I_{DS}$ and $V_{GS}$ dependences will be discussed in Section V. The maximum MR ratio 0.02% was obtained at $V_{GS}$ = 40 V and $I_{DS}$ = 10 mA, which is ~6 times larger than that in our previous paper [8] owing to the reduction of the parasitic resistance in the S/D junctions.

**B. Hanle precession signals measured by the device with $L_{ch}$ = 10 μm**

Hanle precession signals were measured to estimate the spin-related physical parameters in the Si inversion channel, such as the spin lifetime $\tau_S$ and spin diffusion length $\lambda_S$ [15-17,19-21]. For detailed analyses and estimations, Hanle precession signals with several oscillations are needed. Thus, a device with $L_{ch}$ = 10 μm was used for measurements. Figures. 3(a) and (b) show the voltage changes $\Delta V_{SD}$ and $\Delta V_D$ as a function of $H_\perp$, respectively, measured with $V_{GS}$ = 40 V and $I_{DS}$ = 10 mA, where red and blue solid curves are the signals for the parallel (P) and antiparallel (AP) magnetization configurations, respectively. In the 2T measurement setup, the row Hanle precession data include the spin transport signal (2T Hanle signal), local spin extraction signal (narrower three-terminal Hanle (N-3TH) signal), and parasitic magnetoresistance (broader three-terminal Hanle (B-3TH) signals at the S and D electrodes) [22-24]. Since only the 2T Hanle signal changes its sign depending on the P or AP configuration, each signal can be extracted from the row experimental data using the following relations;

$$\Delta V^{3\mathrm{T}} = \frac{\Delta V^{(\mathrm{P})} + \Delta V^{(\mathrm{AP})}}{2}, \tag{1a}$$

$$\Delta V^{2\mathrm{T(P)}} = \frac{\Delta V^{(\mathrm{P})} - \Delta V^{(\mathrm{AP})}}{2}, \tag{1b}$$

$$\Delta V^{2\mathrm{T(AP)}} = \frac{\Delta V^{(\mathrm{AP})} - \Delta V^{(\mathrm{P})}}{2}, \tag{1c}$$

where $\Delta V^{(\mathrm{P})}$ and $\Delta V^{(\mathrm{AP})}$ are the row experimental Hanle precession data obtained in the P and AP configurations, respectively, $\Delta V^{3\mathrm{T}}$ is the three-terminal Hanle signal including both the N-3TH and B-3TH signals, and $\Delta V^{2\mathrm{T(P)}}$ and $\Delta V^{2\mathrm{T(AP)}}$ are the 2T Hanle signals in the P and AP configurations, respectively. In Figs. 3(c) and (d), the red and blue solid curves are $\Delta V^{2\mathrm{T(P)}}$ and $\Delta V^{2\mathrm{T(AP)}}$, respectively, and the green solid curves are $\Delta V^{3\mathrm{T}}$, which were obtained by Eqs. (1a)–(1c). Since the precession signals and the sign reversal depending on the relative magnetization configuration (parallel or antiparallel) were clearly observed in the 2T Hanle signals, the spin transport from the S to D electrodes through the 10-μm-long Si 2D inversion channel was confirmed. We found that the shapes of the 2T Hanle signals in Figs. 3(c) and (d) are identical but the amplitudes in $\Delta V_{\mathrm{D}}$ are almost half of those in $\Delta V_{\mathrm{SD}}$. This probably comes from the current crowding near the left-hand edge of the D electrode: Owing to the current crowding, the electrical potential of the R2 electrode lies between those of the D electrode and the left-hand edge of the $n^{+}$-Si region under the D electrode, leading to the result that the amplitude of the 2T Hanle signal in $V_{\mathrm{D}}$ is almost half of that in $V_{\mathrm{DS}}$ (see Section S1 in Supplemental Material (S.M.) for more detailed explanation [25]).

On the other hand, the 3T Hanle signal $\Delta V_{\mathrm{D}}^{3\mathrm{T}}$ in Fig. 3(d) is a positive narrower bell-shaped curve. This is the N-3TH signal due to the spin extraction of the D electrode, which agrees with our prediction that an Fe/Mg/MgO/$n^{+}$-Si tunnel junction can extract the spins from the $n^+$-Si region, but an Fe/Mg/MgO/$n$-Si Schottky-tunnel junction can not [8]. In contrast, the 3T Hanle signal $\Delta V_{SD}^{3T}$ in Fig. 3(c) is a negative bell-shaped curve, which disagrees with our expectation that both N-3TH and B-3TH signals are positive [23,24]. The reason for this disagreement is unclear at present.

## IV. Analysis of the spin transport signals using the spin drift-diffusion model

### A. Theoretical model for our spin MOSFET structure

In this section, a mathematical 2T Hanle signal expression is derived to analyze the spin transport signals obtained in Section III.B. The difference from our previous paper [8] is that the spin drift effect in the inversion channel and the $n^+$-Si regions in the S/D junctions are taken into account in the model channel structure. As will be described in Section V, the $n^+$-Si regions in the both S and D junctions strongly affect the magnitude of the 2T spin-valve and Hanle signals.

Figure 4(a) shows the side view of the model channel structure in our spin MOSFETs, in which the 2D inversion channel (dark blue region) is formed at the bottom of the 8-nm-thick $p$-Si layer (light blue region) by the back-gate electric field. Electrons flow mainly through the "$n^+$-Si region (yellow region)" and "2D channel region (dark blue region)";

($n^+$-Si region) $\quad -L_S \le y \le 0$ and $L_{ch} \le y \le L_D$,

(2D channel region) $\quad y \le -L_S$, $0 \le y \le L_{ch}$, and $L_{ch} + L_D \le y$.

The device structure is uniform along the $x$ direction, the thicknesses of the $n^+$-Si region and 2D channel region along the $z$ direction are 5 and ~2 nm, respectively, and they are much smaller than the channel length (0.4 or 10 μm) along the $y$ direction. In this channel structure, the distribution of the spin density along the $y$ direction can be descried by the following one-dimensional (1D) spin drift-diffusion equation [15,21,24-26];

$$D_e \frac{d^2}{dy^2} S(y) - v_{\mathrm{d}} \frac{d}{dy} S(y) - \frac{1}{\hat{\tau}_{\mathrm{S}}} S(y) + \frac{P_{\mathrm{S}} I_{\mathrm{DS}}}{q} \delta(y) - \sigma^{\mathrm{P/AP}} \frac{P_{\mathrm{S}} I_{\mathrm{DS}}}{q} \delta(y - L_{\mathrm{ch}}) = 0, \qquad (2)$$

where $S(y)$ [cm$^{-2}$] is the sheet spin density, $D_e$ is the electron diffusion coefficient, $v_{\mathrm{d}} = -\mu_{\mathrm{ch}} F(y)$ is the drift velocity, $\mu_{\mathrm{ch}}$ is the electron mobility, $F(y)$ is the lateral electric field, $\hat{\tau}_{\mathrm{S}} = \tau_{\mathrm{S}} / (1 + i \gamma_e \tau_{\mathrm{S}} H_{\perp})$ is the complex spin lifetime taking into account the spin precession [24], $\tau_{\mathrm{S}}$ is the spin lifetime, $\gamma_e$ (= $1.76 \times 10^7$ s$^{-1}$Oe$^{-1}$) is the gyromagnetic ratio, $\sigma^{\mathrm{P/AP}} = \pm 1$ is the sign parameter expressing the parallel ($\sigma^{\mathrm{P}} = +1$) and an antiparallel ($\sigma^{\mathrm{AP}} = -1$) magnetization configurations of the S and D electrodes, $\delta(y)$ is the Dirac delta function, and $P_{\mathrm{S}}$ is the spin polarization of the injected/extracted electron current from/into the S/D electrode. Note that we use the sheet spin density in Eq. (2) because the effective thickness of the channel can be changed by the gate electric field along the $z$ direction. The first, second, third, fourth, and fifth terms in Eq. (2) express the spin diffusion, spin drift, spin relaxation, spin injection, and spin extraction, respectively. Here, it is assumed that the spin injection and extraction are concentrated at the right-hand edge of the S electrode ($y$ = 0) and the left-hand edge of the D electrode ($y = L_{\mathrm{ch}}$), respectively, because the electron current is concentrated at these points. This assumption is consistent with the fact that the 2T Hanle signals obtained in $V_{\mathrm{DS}}$ and $V_{\mathrm{D}}$ have different amplitudes as mentioned in Fig. 3 (c) and (d) in Section III.B (see also Section S1 in S.M. [25]). In our 2T measurement setup in Fig. 1(d), the lateral electric field $F(y)$ along the $y$ direction is present only in the 2D channel region between the S and D electrodes;

$$F(y)=\begin{cases}-I_{\mathrm{DS}}R_{\mathrm{ch}}/L_{\mathrm{ch}}=-I_{\mathrm{DS}}R_{\mathrm{S}}/W_{\mathrm{ch}} & (0\le y\le L_{\mathrm{ch}})\\ 0 & (y\le 0,\ L_{\mathrm{ch}}\le y)\end{cases}, \tag{3}$$

where $R_{\mathrm{ch}}$ is the channel resistance and $R_{\mathrm{S}}$ (= $R_{\mathrm{ch}}W_{\mathrm{ch}}/L_{\mathrm{ch}}$) is the sheet resistance of the 2D channel. By the continuity of the spin current and the electro-chemical potentials of both up- and down-spins at the four boundaries ($y$ = $-L_{\mathrm{S}}$, 0, $L_{\mathrm{ch}}$, and $L_{\mathrm{ch}}+L_{\mathrm{D}}$), the 2T Hanle signal $\Delta V_{\mathrm{D}}^{\mathrm{2TH(P/AP)}}$ detected by the D electrode are obtained from Eqs. (1)–(3) (the detailed derivation is shown in Section S2 in S.M. [25]);

$$\Delta V_{\mathrm{D}}^{\mathrm{2TH(P/AP)}}=-\sigma^{\mathrm{P/AP}}\,\mathrm{Re}\left[\frac{P_{\mathrm{S}}{}^{2}I_{\mathrm{DS}}}{X}\left(\frac{1}{r_{\mathrm{ch}}^{\mathrm{u}}}+\frac{1}{r_{\mathrm{ch}}^{\mathrm{d}}}\right)\gamma^{\mathrm{d}}\right], \tag{4a}$$

$$X=\left(\frac{1}{r_{\mathrm{NL}}^{(\mathrm{S})}}+\frac{1}{r_{\mathrm{ch}}^{\mathrm{d}}}\right)\left(\frac{1}{r_{\mathrm{NL}}^{(\mathrm{D})}}+\frac{1}{r_{\mathrm{ch}}^{\mathrm{u}}}\right)-\left(\frac{1}{r_{\mathrm{NL}}^{(\mathrm{S})}}-\frac{1}{r_{\mathrm{ch}}^{\mathrm{u}}}\right)\left(\frac{1}{r_{\mathrm{NL}}^{(\mathrm{D})}}-\frac{1}{r_{\mathrm{ch}}^{\mathrm{d}}}\right)\gamma^{\mathrm{u}}\gamma^{\mathrm{d}}, \tag{4b}$$

$$\gamma^{\mathrm{u}}=\exp\left(-\frac{L_{\mathrm{ch}}}{\lambda_{\mathrm{ch}}^{\mathrm{u}}}\right),\ \gamma^{\mathrm{d}}=\exp\left(-\frac{L_{\mathrm{ch}}}{\lambda_{\mathrm{ch}}^{\mathrm{d}}}\right), \tag{4c}$$

$$\frac{1}{\lambda_{\mathrm{ch}}^{\mathrm{d}}}=-\frac{1}{2\Lambda}+\sqrt{\left(\frac{1}{2\Lambda}\right)^{2}+\frac{1}{\lambda_{\mathrm{ch}}{}^{2}}},\ \frac{1}{\lambda_{\mathrm{ch}}^{\mathrm{u}}}=\frac{1}{2\Lambda}+\sqrt{\left(\frac{1}{2\Lambda}\right)^{2}+\frac{1}{\lambda_{\mathrm{ch}}{}^{2}}}, \tag{4d}$$

$$\frac{1}{\Lambda}=\frac{v_{\mathrm{d}}}{D_{e}^{\mathrm{ch}}}=\frac{\mu_{\mathrm{ch}}}{D_{e}^{\mathrm{ch}}}\frac{I_{\mathrm{DS}}R_{\mathrm{S}}}{W_{\mathrm{ch}}}, \tag{4e}$$

where $\Lambda$ is the effective diffusion length of electrons against the drift, $\lambda_{\mathrm{ch}}^{\mathrm{d}}$ and $\lambda_{\mathrm{ch}}^{\mathrm{u}}$ are the down-stream and up-stream spin drift diffusion lengths [14] in the 2D channel region, respectively, $\lambda_{\mathrm{ch}}=\sqrt{D_{e}^{\mathrm{ch}}\tau_{\mathrm{S}}^{\mathrm{ch}}}$ ($\lambda_{\mathrm{n}}=\sqrt{D_{e}^{\mathrm{n}}\tau_{\mathrm{S}}^{\mathrm{n}}}$), $D_{e}^{\mathrm{ch}}$ ($D_{e}^{\mathrm{n}}$), and $\tau_{\mathrm{S}}^{\mathrm{ch}}$ ($\tau_{\mathrm{S}}^{\mathrm{n}}$) are the intrinsic spin diffusion length without the lateral electric field along the $y$ direction, electron diffusion coefficient, and spin lifetime in the 2D channel region ($n^{+}$-Si region), respectively, $r_{\mathrm{ch}}^{\mathrm{d}}$ and $r_{\mathrm{ch}}^{\mathrm{u}}$ are the down-stream and up-stream spin resistances, respectively, and $r_{\rm NL}^{\rm (S)}$ and $r_{\rm NL}^{\rm (D)}$ are the effective spin resistances of the left-hand side of the spin injection point ($y \le 0$) and the right-hand side of the spin detection point ($L_{\rm ch} \le y$), respectively. Here, $\lambda_{\rm ch}^{\rm d}$ is the effective spin diffusion length for the spins transported from the S to D electrodes, whereas $\lambda_{\rm ch}^{\rm u}$ is that for the spins transported from the D to S electrodes. On the other hand, $r_{\rm ch}^{\rm d}$ and $r_{\rm ch}^{\rm u}$ are regarded as the effective spin resistances of the local region ($0 \le y \le L_{\rm ch}$) viewed from the S and D sides, respectively. Figure 4(b) shows an equivalent circuit of Fig. 4(a) expressed by the spin resistances, where $r_{\rm ch}^{\rm d}$ and $r_{\rm ch}^{\rm u}$ are the spin resistances for the spin transport along the $+y$ and $-y$ directions, respectively. These spin resistances are very helpful to analyze and understand the spin transport, as will be shown in Section V. Here, $r_{\rm ch}^{\rm d}$, $r_{\rm ch}^{\rm u}$, $r_{\rm NL}^{\rm (S)}$ and $r_{\rm NL}^{\rm (D)}$ are expressed by;

$$r_{\rm ch}^{\rm d} = \frac{R_{\rm S}\lambda_{\rm ch}^{\rm u}}{W_{\rm ch}} \tag{5a}$$

$$r_{\rm ch}^{\rm u} = \frac{R_{\rm S}\lambda_{\rm ch}^{\rm d}}{W_{\rm ch}} \tag{5b}$$

$$r_{\rm NL}^{\rm (S)} = \frac{(r_{\rm ch}+r_{\rm n})+(r_{\rm ch}-r_{\rm n})\exp\left(-2L_{\rm S}/\lambda_{\rm n}\right)}{(r_{\rm ch}+r_{\rm n})-(r_{\rm ch}-r_{\rm n})\exp\left(-2L_{\rm S}/\lambda_{\rm n}\right)} r_{\rm n}, \tag{5c}$$

$$r_{\rm NL}^{\rm (D)} = \frac{(r_{\rm ch}+r_{\rm n})+(r_{\rm ch}-r_{\rm n})\exp\left(-2L_{\rm D}/\lambda_{\rm n}\right)}{(r_{\rm ch}+r_{\rm n})-(r_{\rm ch}-r_{\rm n})\exp\left(-2L_{\rm D}/\lambda_{\rm n}\right)} r_{\rm n}, \tag{5d}$$

where $r_{\rm ch} = R_{\rm S}\lambda_{\rm ch}/W_{\rm ch}$ and $r_{\rm n} = \rho_{\rm n}\lambda_{\rm n}/t_{\rm n}W_{\rm ch}$ are the intrinsic (without spin drift) spin resistances of the 2D channel region and $n^+$-Si region, respectively, and $\rho_{\rm n}$ is the electrical resistivity of the $n^+$-Si region. We should emphasize here that up(down)-stream spin resistance $r_{\rm ch}^{\rm u(d)}$ is proportional to down(up)-stream spin drift-diffusion length $\lambda_{\rm ch}^{\rm d(u)}$ as shown in Eqs. (5a) and (5b). This is because the spin resistance of a non-magnetic material is inversely proportional to its effective spin diffusion length when $\tau_S$ is unchanged, and up- and down-stream spin drift-diffusion lengths are inversely proportional to each other ($\lambda_{ch}^{u}\lambda_{ch}^{d} = (\lambda_{ch})^2$) (see section S3 in S.M. for the detail [25]). The magnitude of spin-valve signal $\Delta V^{2T}$ can be calculated from Eq. (4a), since it is the difference between $\Delta V_D^{2TH(AP)}$ and $\Delta V_D^{2TH(P)}$ at $H_\perp = 0$;

$$\begin{aligned}\Delta V^{2T} &= \Delta V_D^{2TH(AP)}\Big|_{H_\perp=0} - \Delta V_D^{2TH(P)}\Big|_{H_\perp=0} \\ &= \frac{2P_S^{\,2} I_{DS}}{X}\left(\frac{1}{r_{ch}^{u}}+\frac{1}{r_{ch}^{d}}\right)\gamma^{d}\Bigg|_{H_\perp=0}\end{aligned} \qquad (6)$$

The physical interpretation of Eqs. (4a)–(4e) is as follows. When $|F(y)|$ is substantially large by a large $I_{DS}$ (>0), the spin polarization decays with $\lambda_{ch}^{d}$ (> $\lambda_{ch}$) from the S to D electrodes, while the spin polarization decays with $\lambda_{ch}^{u}$ (< $\lambda_{ch}$) from the D to S electrodes. Thus, the effective spin diffusion length can be changed by the lateral electric field along the $y$ direction. When the spin drift dominates the spin transport, $\lambda_{ch}^{d}$ approaches the spin drift length $v_d\tau_S$ and $\lambda_{ch}^{u}$ approaches 0. A remarkable feature is that $r_{ch}^{d}$ decreases and $r_{ch}^{u}$ increases as $I_{DS}$ increases, because $\lambda_{ch}^{d}$ increases and $\lambda_{ch}^{u}$ decreases as $I_{DS}$ increases.

To analyze the spin transport signals using Eqs. (4a) and (6), some electrical parameters, such as $R_S$ and $\rho_n$, were estimated using two Hall-bar-type MOSFET devices; type-I has the same channel properties as those in the spin MOSFET of Fig. 1 and type-II has a highly-phosphorus-doped $n^+$-Si channel that was formed by the same procedure as that for the $n^+$-Si regions in the spin MOSFET. Figure 5(a) shows a schematic illustration of the type-I device where the channel length and width are 460 μm and 90 μm, respectively, and the Si channel thickness $t_{Si}$ is 8 nm. The measurement setup is also shown in the same figure, in which a constant bias current $I_{DS}$ and a constant source-gate voltage $V_{GS}$ were applied, and the longitudinal voltage $V_L$ and transverse voltages $V_T$ were measured while a magnetic field was applied perpendicular to the substrate plane. From the measurement results, the following values were estimated; $R_S$ = 8481, 4604, and 3370 Ω for $V_{GS}$ = 40, 60, and 80 V, respectively. On the other hand, the sheet electron density $N_S$, the electron mobility $\mu_{ch}$, the diffusion coefficient $D_e^{ch}$, and the electron momentum lifetime $\tau^{ch}$ in the inversion channel were also estimated using the same procedure as our previous paper [8] (see Section S4 in S.M. for the detail [25]). In Fig. 5(b), the blue-filled circles and pink-filled squares are $N_S$ and $R_S$, respectively, as a function of $V_{GS}$. We confirmed that the slope of $N_S$ is consistent with the capacitance of the back-gate dielectric. In Fig. 5(c), the green open diamonds are $\mu_{ch}$ as a function of the effective gate electric field $E_{eff} = q/\varepsilon_{Si}\left(N_S/2 + N_A t_{Si}\right)$ [27], where $q$ is the elementary charge and $\varepsilon_{Si}$ is the permittivity of Si. In Fig. 5(d), a blue solid curve is $D_e^{ch}$ as a function of $E_{eff}$, which was estimated by a self-consistent calculation.

Using the same measurement setup with $V_{GS}$ = 0 V, $\rho_n$ = 0.89 mΩcm was estimated from the type-II device. Then, $D_e^{n}$ = 3.5 cm$^2$/s was obtained by the Einstein's relation [28].

## B. Procedure of the analysis

Using the functions Eqs. (4a) and Eq. (6), we analyze the experimental 2T Hanle signals and spin-valve signals. In Figs. 3(c) and (d), whereas the amplitudes in $\Delta V_{\mathrm{D}}$ are almost half of those in $\Delta V_{\mathrm{DS}}$, clear oscillations in the higher magnetic fields are seen in $V_{\mathrm{D}}$ because of its high signal to noise ratio. Since the fitting function Eq. (4a) for the experimental 2T Hanle signals is nonlinear, clear oscillation signals with larger number of peaks are necessary for accurate estimation. Thus, we use the 2T Hanle signals in $V_{\mathrm{D}}$ to estimate $\mu_{\mathrm{ch}}$, $D_{\mathrm{e}}^{\mathrm{ch}}$, $\tau_{\mathrm{S}}^{\mathrm{ch}}$, and $\tau_{\mathrm{S}}^{\mathrm{n}}$. Since the actual signal magnitude is necessary to estimate $P_{\mathrm{S}}$, we estimate $P_{\mathrm{S}}$ by fitting Eq. (6) to the $\Delta V^{\mathrm{2T}}$ values measured with various $I_{\mathrm{DS}}$, while $\mu_{\mathrm{ch}}$, $D_{e}^{\mathrm{ch}}$, $\tau_{\mathrm{S}}^{\mathrm{ch}}$, and $\tau_{\mathrm{S}}^{\mathrm{n}}$ are set at the values estimated from the 2T Hanle signals. Then, we obtain $\lambda_{\mathrm{n}}$, $\lambda_{\mathrm{ch}}$, $\lambda_{\mathrm{ch}}^{\mathrm{d}}$ and $\lambda_{\mathrm{ch}}^{\mathrm{u}}$ values using the above-mentioned parameters and discuss how $\Delta V^{\mathrm{2T}}$ changes depending on both $I_{\mathrm{DS}}$ and $V_{\mathrm{GS}}$.

**C. Analysis of the Hanle signals**

Figure 3(e) shows the Hanle precession signals obtained in the spin MOSFET with $L_{\mathrm{ch}}$ = 10 μm, where red solid curves are the experimental 2T Hanle signals $\Delta V_{\mathrm{D}}^{\mathrm{2TH(P)}}$ measured at $V_{\mathrm{GS}}$ = 40 V with $I_{\mathrm{DS}}$ = 10, 8, and 5 mA, respectively. In spite of the same $V_{\mathrm{GS}}$, the period of the oscillation inceases with increasing $I_{\mathrm{DS}}$. Since $N_{\mathrm{S}}$ in the channel is the same for all the $I_{\mathrm{DS}}$ cases, the increase in the period is caused by the increase in $v_{\mathrm{d}}$. Figure 3(f) shows the 2T Hanle signals obtained in the same device, where red solid curves are the experimental 2T Hanle signals $\Delta V_{\mathrm{D}}^{\mathrm{2TH(P)}}$ measured at $I_{\mathrm{DS}}$ = 10 mA with $V_{\mathrm{GS}}$ = 40, 60, and 80 V. In spite of the same $I_{\mathrm{DS}}$, the period of the oscillation deceases with increasing $V_{\mathrm{GS}}$. This is mainly caused by the increase in $N_{\mathrm{S}}$, since $\mu_{\mathrm{ch}}$ and $D_{e}^{\mathrm{ch}}$ are almost unchanged with increasing $V_{\mathrm{GS}}$ from 40 to 80 V, as shown in Figs. 5(c) and (d), while $N_{\mathrm{S}}$ is doubled as shown in Fig. 5(b).

Using $R_S$ and $D_e^n$ obtained in Section IV.A, the experimental signals in Figs. 3(e) and (f) were fitted by Eq. (4a) with fitting parameters $P_S$, $\mu_{ch}$, $D_e^{ch}$, $\tau_S^{ch}$, and $\tau_S^n$. Among these fitting parameters, $\mu_{ch}$, $D_e^{ch}$, and $\tau_S^{ch}$ are assumed to depend only on $V_{GS}$ (indeendent of $I_{DS}$), while $\tau_S^n$ is assumed to be independent of both $V_{GS}$ and $I_{DS}$. In Fig. 3(e), all the black dashed curves are fitting curves with the same parameter values: $\mu_{ch}$ = 239 cm$^2$/Vs, $\mu_{ch}$, $D_e^{ch}$ = 11.3 cm$^2$/s, $\tau_S^{ch}$ = 0.82 ns, and $\tau_S^n$ = 0.6 ns. In Fig. 3(f), all the black dashed curves are fitting curves and the following values were estimated: $\mu_{ch}$ = 253 cm$^2$/Vs, $D_e^{ch}$ = 11.0 cm$^2$/s, $\tau_S^{ch}$ = 0.84 ns for $V_{GS}$ = 60 V, and $\mu_{ch}$ = 241 cm$^2$/Vs, $D_e^{ch}$ = 10.8 cm$^2$/s, $\tau_S^{ch}$ = 0.77 ns for $V_{GS}$ = 80 V. In both Figs. 3(e) and (f), the experimental signals are almost perfectly reproduced by Eq. (4a). Therefore, the analytical function Eq. (4a) accurately expresses the spin transport phenomena in our spin MOSFETs.

To investigate further, the parameters $\mu_{ch}$, $D_e^{ch}$, and $\tau_S^{ch}$ estimated by the type-I device in Fig. 5(a) were compared with those estimated from the fittings to the 2T Hanle signals. In Fig. 5(c), $\mu_{ch}$ estimated from the fittings are plotted by red open squares, which agree well with the green diamonds estimated from the Hall measurements. In Fig. 5(d), $D_e^{ch}$ estimated from the fittings are plotted by red open squares, which reasonably agree with the blue curve obtained by our self-consistent calculation (see section S4 in S.M. [25]). These results strongly support the accuracy of our analysis with the 2T Hanle signals. On the other hand, the spin lifetime $\tau_S^{ch}$ values estimated from the Hanle measurements are plotted by the red open squares in Fig. 5(e), where the $\tau^{\text{ch}}$ values multiplied by 25000 are also plotted by a blue curve obtained by our self-consistent calculation (see Section S4 in S.M. [25]). Since the red open squares overlap the blue curve, the spin-flip probability (spin-flip rate per one momentum scattering event) is expressed by $\tau^{\text{ch}} / \tau_{\text{S}}^{\text{ch}} \sim 1/25000$ in the spin MOSFET examined in this study. This spin-flip probability value is somehow smaller than that in our previous spin MOSFET ($\tau^{\text{ch}} / \tau_{\text{S}}^{\text{ch}} \sim 1/14000$) with a *n*-type Si channel having a phosphorus donor concentration of $10^{17}$ $\text{cm}^{-3}$ [8]. This suggests that the strength of spin-orbit coupling (SOC) is reduced in the present spin MOSFETs (with $N_{\text{A}} \sim 1\times10^{5}$ $\text{cm}^{-3}$ in the channel) examined in this study due to the lack of ionized phosphorus donors in the channel. Using the $D_e^{\text{ch}}$ and $\tau_{\text{S}}^{\text{ch}}$ values estimated from the 2T Hanle signals, the following $\lambda_{\text{ch}}$ values were obtained: 0.96 μm for $V_{\text{GS}}$ = 40 V, 0.96 μm for $V_{\text{GS}}$ = 60 V, and 0.91 μm for $V_{\text{GS}}$ = 80 V. These values will be used later in Sections IV.D and V. Also, $\lambda_{\text{n}}$ = 0.46 μm was estimated from the $D_e^{\text{n}}$ and $\tau_{\text{S}}^{\text{n}}$ values. A notable feature is that the clear 2T Hanle signals were obtained in the spin MOSFET with $L_{\text{ch}}$ = 10 μm, even though $\lambda_{\text{ch}}$ was estimated to be only ~1 μm for all $V_{\text{GS}}$. This is caused by the enhancement of $\lambda_{\text{ch}}^{\text{d}}$ by the spin drift [14]. Figure 6(a) shows the spin drift-diffusion lengths calculated using Eq. (4d), where red, green, and blue solid curves are $\lambda_{\text{ch}}^{\text{d}}$ at $V_{\text{GS}}$ = 40, 60, and 80 V, respectively, and the black solid line is $\lambda_{\text{ch}}$. In the same figure, the spin drift length ($v_{\text{d}}\tau_{\text{S}}^{\text{ch}}$) at $V_{\text{GS}}$ = 40, 60, and 80 V are also shown by red, green, and blue dashed lines, respectively, which indicate that the spin drift dominates the spin transport when $I_{\text{DS}}$ > 5 mA. As $I_{\text{DS}}$ increases, $\lambda_{\text{ch}}^{\text{d}}$ increases and exceeds 10 μm when $I_{DS}$ > 10 mA at $V_{GS}$ = 40 V. Therefore, it is reasonable that the 2T Hanle signals were obtained in the spin MOSFET with $L_{ch}$ = 10 μm. From the analysis of the 2T Hanle signal measured at $I_{DS}$ = 10 mA and $V_{GS}$ = 40 V for the spin MOSFET with $L_{ch}$ = 0.4 μm, 96% of spins injected in the 2D inversion channel at $y$ = 0 reaches $y$ = $L_{ch}$, i.e., almost all the spins are conserved during the transport through the channel.

Note that Eq. (4a) becomes identical with the conventional function [15-17] by replacing $r_{NL}^{(S)}$ with $r_{ch}^{u}$ and $r_{NL}^{(D)}$ with $r_{ch}^{d}$, but the conventional function can *not* reproduce our 2T Hanle signals shown in Figs. 3(e) and (f) (see Section S5 in S.M.). Thus, our channel model structure in Fig. 4(a) is appropriate for the analysis of the 2T Hale signals.

**D. Analysis of the spin-valve signals**

In the following analysis, $\mu_{ch}$, $D_e^{ch}$, $\tau_S^{ch}$, and $\tau_S^{n}$ values are set to the values estimated from the 2T Hanle signals in Figs. 3(e) and (f) and only $P_S$ is the fitting parameter. Figure 6(b) shows the bias dependences of $\Delta V^{2T}$ measured at $V_{GS}$ = 40 V, where red and purple filled circles are experimental data of the spin MOSFETs with $L_{ch}$ = 0.4 (Fig. 2(c)) and 10 μm, respectively, and red and purple solid curves are the theoretical calculations using Eq. (6) with $P_S$ = 7.3%. The good agreement means that the superlinear bias dependence of the spin-valve signals comes from the spin drift, since a linear bias dependence would be obtained without spin drift. As shown in Fig. 6(b), a spin-valve signal was not obtained at $I_{DS}$ = 2 mA in the spin MOSFET with $L_{ch}$ = 10 μm. This is consistent with the fact that $\lambda_{ch}^{d}$ was estimated to be only ~2 μm at $I_{DS}$ = 2 mA as shown in Fig. 6(a), which leads to the situation that the spins injected from the S electrode can not reach the D electrode. Figure 6(c) shows the bias dependences of $\Delta V^{2T}$ with various $V_{GS}$ (= 40, 60, and 80 V), where red, green, and blue filled circles are experimental data (Fig. 2(d)) obtained for the spin MOSFET with $L_{ch}$ = 0.4 μm, respectively, and red, green, and blue solid curves are theoretical calculations using Eq. (6) with $P_S$ = 7.3%, 6.5 %, and 5.6%, respectively. The bias dependences are well reproduced by our model calculation in all the $V_{GS}$ values examined in this study. Thus, $P_S$ decreases with increasing $V_{GS}$. The same feature was also obtained in our previous device [8], but the origin is unclear at present.

Figure 6(d) shows the MR ratio of the spin-MOSFET device with $L_{ch}$ = 0.4 μm. The MR ratio increases with increasing $I_{DS}$ and with decreasing $V_{GS}$. These features mainly come from the increase in $\lambda_{ch}^{d}$ and $P_S$, since $\lambda_{ch}^{d}$ increases with increasing $I_{DS}$ and with decreasing $V_{GS}$, and $P_S$ increases with decreasing $V_{GS}$.

**V. Discussion**

We have shown that the almost all the experimental signals can be explained by Eqs. (4a) and (6) in Section IV. In this section, to further investigate how the spin drift affects the magnitude of the spin-valve signal $\Delta V^{2T}$, we perform the calculation of the spin accumulation voltage and spin current distribution in the Si channel structure consisting of the $n^+$-Si regions and inversion channel. Then, we introduce "input spin resistance $r^{\text{input}}$" and "output spin resistance $r^{\text{output}}$", which are the effective spin resistances of the whole spin MOSFET viewed from the S and D electrodes, respectively. The analysis with $r^{\text{input}}$ and $r^{\text{output}}$ allows us to easily capture how the spin injection and detection efficiencies are changed by the spin drift; smaller $r^{\text{input}}$ and larger $r^{\text{output}}$ lead to higher spin injection and detection efficiencies, respectively. We find that the spin drift effectively decreases $r^{\text{input}}$, but it hardly increases $r^{\text{output}}$ in the present spin MOSFET structure. Next, using the experimental results, we analyze $r^{\text{input}}$ and $r^{\text{output}}$ and find that a part of the injected spin current does not contribute to the spin-valve signal, i.e., not all of the injected spin current are transported from the S electrode and accumulated under the D electrode, but the injected spin current is partially "wasted" in the $n^+$-Si regions at the S and D junctions. Based on these results, we clearly show that the design of these $n^+$-Si regions is a key to further improve the MR ratio.

**A. Spin distribution in the spin MOSFET with different $I_{\text{DS}}$**

In our spin MOSFETs, the electron density and the density of states (DOS) are not uniform in the entire channel structure consisting of the $n^+$-Si regions and inversion channel. Thus, the sheet spin density $S(y)$ becomes discontinuous at the boundaries and is not suitable to describe the spin distribution in the channel. For this case, it is more suitable to use the spin accumulation $\Delta E = (E_+ - E_-)/2$ that is half of the difference in the electro-chemical potential between up-spin ($E_+$) and down-spin ($E_-$) electrons (see Section S2 in S.M. for details [25]).

Figures 7(a) and (b) show normalized spin accumulation $\Delta E / qI_{\text{DS}}$ [Ω] calculated by Eq. (2) for the device with $I_{\text{DS}}$ = 2 and 10 mA, respectively, where $L_{\text{ch}}$ = 10 μm, $V_{\text{GS}}$ = 40 V, and red and blue solid curves represent $\Delta E / qI_{\text{DS}}$ in P and AP configurations, respectively. In the calculation, we assume $P_{\text{S}}$ = 1 for simplicity. Figures 7(c) and (d) show close-up views of (a) and (b) at around $y$ = 10 μm (near the D electrode), respectively. In each figure, the black dashed curve is $\Delta E / qI_{\text{DS}}$ injected from the S electrode, which was calculated without the fifth term of Eq. (2), whereas the black dotted curve is that extracted into the D electrode, which was calculated without the fourth term of Eq. (2). As mentioned in Section IV, when $V_{GS}$ = 40 V and $I_{DS}$ = 2 mA, spins injected from the S electrode do not reach the D electrode because $\lambda_{ch}^{d}$ is only ~ 2 μm. This can be clearly seen in Fig. 7(a); the black dashed curve exponentially decays from the S electrode toward the D electrode and becomes almost zero at around $y$ = 10 μm. In this situation, the spin density near the D electrode ($y$ ~ 10 μm) is determined only by the spins extracted from the D electrode, which is represented by the dotted curve. Since there is no difference between the P and AP configurations at the D electrode ($y$ = 10 μm) as seen in Fig. 7(c), no spin-valve signal is obtained.

On the other hand, since $\lambda_{ch}^{d}$ is ~10 μm at $I_{DS}$ = 10 mA, spins injected from the S electrode reach the D electrode as shown in Fig. 7(b); the dashed curve has a non-zero value at $y$ = 10 μm. In this situation, a spin-valve signal $\Delta V^{2T}$ is obtained, since there is a difference in $\Delta E / qI_{DS}$ between the P and AP configurations at the D electrode ($y$ = 10 μm), which is indicated by a double-headed black arrow in Figs. 7(b) and (d).

Other notable features are (i) normalized spin accumulation $\Delta E / qI_{DS}$ near the S electrode ($y$ ~ 0 μm) in Fig. 7(b) is smaller than that in Fig. 7(a) and (ii) $\Delta E / qI_{DS}$ near the D electrode ($y$ ~ 10 μm) in Fig. 7(a) is almost the same as that in Fig. 7(b). These are strongly related to the spin injection and detection efficiencies, as will be explained in the following subsections.

**B. Expression of the spin-valve signals with input and output spin resistances**

When the spin drift dominates the spin transport in the inversion channel, the spin back-flow from the D to S electrodes is negligible ($\lambda_{\text{ch}}^{\text{u}} << L_{\text{ch}}$ and $\gamma^{u} \sim 0$) and $r_{\text{ch}}^{\text{u}}$ becomes much larger than $r_{\text{ch}}^{\text{d}}$ ($r_{\text{ch}}^{\text{d}} << r_{\text{ch}}^{\text{u}}$). In this situation, Eq. (6) can be written by the following simple form;

$$\Delta V^{2\text{T}} \sim 2 \times P_{\text{S}} I_{\text{DS}} \times \frac{r^{\text{input}}}{r_{\text{ch}}^{\text{d}}} \times \gamma^{\text{d}} \times P_{\text{S}} r^{\text{output}} . \tag{7}$$

Here, $r^{\text{input}}$ and $r^{\text{output}}$ are the input spin resistance and output spin resistance, respectively, which are defined by the parallel resistances of ($r_{\text{NL}}^{(\text{S})}$ and $r_{\text{ch}}^{\text{d}}$) and ($r_{\text{NL}}^{(\text{D})}$ and $r_{\text{ch}}^{\text{u}}$), respectively;

$$r^{\text{input}} = \left( \frac{1}{r_{\text{NL}}^{(\text{S})}} + \frac{1}{r_{\text{ch}}^{\text{d}}} \right)^{-1} , \tag{8a}$$

$$r^{\text{output}} = \left( \frac{1}{r_{\text{NL}}^{(\text{D})}} + \frac{1}{r_{\text{ch}}^{\text{u}}} \right)^{-1} , \tag{8b}$$

where $r_{\text{ch}}^{\text{d}}$, $r_{\text{ch}}^{\text{u}}$, $r_{\text{NL}}^{(\text{S})}$, and $r_{\text{NL}}^{(\text{D})}$ are the spin resistances defined by Eqs. (5a)–(5d). It will be shown later that the term $P_{\text{S}} I_{\text{DS}} \times r^{\text{input}} / r_{\text{ch}}^{\text{d}}$ in Eq. (7) is related to the spin injection efficiency, $\gamma^{\text{d}}$ ($= \exp(-L_{\text{ch}} / \lambda_{\text{ch}}^{\text{d}})$) is related to the spin transport efficiency, and $P_{\text{S}} r^{\text{output}}$ is related to the spin detection efficiency. To understand the physical meaning of Eq. (7), equivalent circuits consisting of these spin resistances is helpful for discussion.

Figure 8(a) shows an equivalent circuit consisting of the spin resistances around the S electrode, where $I_{\text{S}}^{\text{inj}} = P_{\text{S}} I_{\text{DS}}$ is the spin current injected from the S electrode, $I_S^{\mathrm{local}}$ is the spin current that flows into the local region ( $y > 0$ ), and $I_S^{\mathrm{nonlocal(S)}}$ is the spin current that flows into the nonlocal region ( $y < 0$ ). From the spin current continuity, $I_S^{\mathrm{inj}} = I_S^{\mathrm{local}} + I_S^{\mathrm{nonlocal(S)}}$ holds at $y$ = 0. Thus, $r^{\mathrm{input}}$ is the parallel resistance of $r_{\mathrm{ch}}^{\mathrm{d}}$ and $r_{\mathrm{NL}}^{(\mathrm{S})}$, as defined in Eq. (8a). The spin-valve signal is produced by $I_S^{\mathrm{local}}$ that is the spin current transported to the D electrode, whereas $I_S^{\mathrm{nonlocal}}$ diffuses toward the nonlocal region and does not contribute to the spin-valve signal. Since the ratio of $I_S^{\mathrm{local}}$ to $I_S^{\mathrm{nonlocal}}$ is determined by the ratio $r_{\mathrm{NL}}^{(\mathrm{S})}$ to $r_{\mathrm{ch}}^{\mathrm{d}}$, we can estimate the local spin current ratio by,

$$\frac{I_S^{\mathrm{local}}}{I_S^{\mathrm{inj}}} = \frac{I_S^{\mathrm{local}}}{I_S^{\mathrm{local}} + I_S^{\mathrm{nonlocal}}} = \frac{1/r_{\mathrm{ch}}^{\mathrm{d}}}{1/r_{\mathrm{ch}}^{\mathrm{d}} + 1/r_{\mathrm{NL}}^{(\mathrm{S})}} = \frac{r^{\mathrm{input}}}{r_{\mathrm{ch}}^{\mathrm{d}}}. \tag{9}$$

Thus, the term $P_S I_{\mathrm{DS}} \times r^{\mathrm{input}} / r_{\mathrm{ch}}^{\mathrm{d}}$ in Eq. (7) is nothing but $I_S^{\mathrm{local}}$, i.e., the effective spin current that is injected into the 2D inversion channel. From the definition, the ratio $r^{\mathrm{input}} / r_{\mathrm{ch}}^{\mathrm{d}}$ $(= r_{\mathrm{NL}}^{(\mathrm{S})} /(r_{\mathrm{ch}}^{\mathrm{d}} + r_{\mathrm{NL}}^{(\mathrm{S})}))$ takes a value between 0 – 1 and $r^{\mathrm{input}} / r_{\mathrm{ch}}^{\mathrm{d}} \sim 1$ when $r_{\mathrm{ch}}^{\mathrm{d}} << r_{\mathrm{NL}}^{(\mathrm{S})}$. Therefore, lower $r_{\mathrm{ch}}^{\mathrm{d}}$ is needed to obtain a higher $I_S^{\mathrm{local}} / I_S^{\mathrm{inj}}$ ratio.

It is noteworthy that $r^{\mathrm{input}}$ determines the conductivity matching condition for the spin injection at the S junction [29-32]. Based on refs. [29-32], the effective spin injection polarization is expressed by $P_S^* = P_S r_B^{(\mathrm{S})} /(r_B^{(\mathrm{S})} + r^{\mathrm{input}})$, where $r_B^{(\mathrm{S})}$ is the tunnel resistance of the S junction (see Section S6 in S.M. for the detail [25]). To obtain higher MR ratio, it is necessary to insert a tunnel barrier with $r_B^{(\mathrm{S})} \sim r^{\mathrm{input}}$ between the FM and Si, because $P_S^*$ decreases when $r_B^{(\mathrm{S})} << r^{\mathrm{input}}$ and the parasitic resistance increases when $r_B^{(\mathrm{S})} >> r^{\mathrm{input}}$. Thus, lower $r^{\mathrm{input}}$ leads to lower $r_B^{(\mathrm{S})}$.

From Eq. (8a), lower $r^{\mathrm{input}}$ is realized by smaller $r_{\mathrm{ch}}^{\mathrm{d}}$ and/or smaller $r_{\mathrm{NL}}^{(\mathrm{S})}$. Since $r_{\mathrm{ch}}^{\mathrm{d}} << r_{\mathrm{NL}}^{(\mathrm{S})}$ is needed to obtain large $I_{\mathrm{S}}^{\mathrm{local}} / I_{\mathrm{S}}^{\mathrm{inj}}$ ratio, smaller $r_{\mathrm{ch}}^{\mathrm{d}}$ is indispensable to reduce the parasitic resistance.

Figure 8(b) shows an equivalent circuit consisting of the spin resistances around the D electrode, in which $I_{\mathrm{S}}^{\mathrm{trans}} = I_{\mathrm{S}}^{\mathrm{local}} \gamma^{\mathrm{d}}$ is the spin current transported from the S electrode through the channel, $I_{\mathrm{S}}^{\mathrm{nonlocal(D)}}$ is the spin current that flows into the nonlocal region ($y > L_{\mathrm{ch}}$), and $I_{\mathrm{S}}^{\mathrm{reflect}}$ is the spin current that flows back to the local region ($y < L_{\mathrm{ch}}$). Since the tunnel resistance $r_{\mathrm{B}}^{(\mathrm{D})}$ of the D junction is high enough in our spin MOSFET, the spin current that flows from the $n^{+}$-Si region to the D electrode is neglected (see Section S6 in S.M. for the detail [25]). From the spin current continuity, $I_{\mathrm{S}}^{\mathrm{trans}} = I_{\mathrm{S}}^{\mathrm{nonlocal(D)}} + I_{\mathrm{S}}^{\mathrm{reflect}}$ holds at $y = L_{\mathrm{ch}}$. Thus, $r^{\mathrm{output}}$ is expressed by the parallel resistance of $r_{\mathrm{ch}}^{\mathrm{u}}$ and $r_{\mathrm{NL}}^{(\mathrm{D})}$, as defined in Eq. (8b). Since Eq. (7) can be written as $\Delta V^{\mathrm{2T}} \sim 2 \times I_{\mathrm{S}}^{\mathrm{trans}} \times P_{\mathrm{S}} r^{\mathrm{output}}$, larger $r^{\mathrm{output}}$ is needed to obtain lager $\Delta V^{\mathrm{2T}}$.

Note that almost ideal spin transport efficiency ($\gamma^{\mathrm{d}} = 0.96$) was achieved by the spin drift in our spin MOSFET under the condition of $L_{\mathrm{ch}}$ = 0.4 μm, $I_{\mathrm{DS}}$ = 10 mA, and $V_{\mathrm{GS}}$ = 40 V. Thus, to obtain higher MR ratios, spin injection and detection efficiencies must be improved.

### C. Spin injection efficiency of the S junction

In the previous section V.B, we showed that the decrease of $r_{\mathrm{ch}}^{\mathrm{d}}$ leads to both a higher $I_{\mathrm{S}}^{\mathrm{local}} / I_{\mathrm{S}}^{\mathrm{inj}}$ $(= r^{\mathrm{input}} / r_{\mathrm{ch}}^{\mathrm{d}})$ ratio and a higher MR ratio. As the spin drift increases ($v_d$ and $I_{DS}$ increases), $\lambda_{ch}^{u}$ decrease, and thus $r_{ch}^{d}$ decreases. When $v_d$ is high enough ($v_d \to \infty$) by the spin drift, $r_{ch}^{d}$ becomes almost zero ($r_{ch}^{d} \to 0$). Thus, increasing the spin drift is very effective to realize a high $P_S$ with a lower tunnel barrier resistance [14].

To analyze the $I_S^{local} / I_S^{inj}$ ratio in our spin MOSFETs, spin resistance values at $V_{GS}$ = 40 V with $I_{DS}$ = 2 and 10 mA were calculated using Eqs. (5a)–(5d), (8a), and (8b), which are summarized in Table 1. When $I_{DS}$ = 2 mA, $I_S^{local} / I_S^{inj}$ is only 21%. Thus, even if $P_S$ = 100% is realized, only 21% of the spins is injected into the inversion channel and the rest (79%) is wasted in the nonlocal region. To increase $I_S^{local} / I_S^{inj}$, the increase of $r_{NL}^{(S)}$ and decrease of $r_{ch}^{d}$ are needed. As shown in Table 1, when $I_{DS}$ is increased from 2 mA to 10 mA, $r_{ch}^{d}$ is decreased from 18.5 Ω to 4.4 Ω, which is caused by the spin drift . When $I_{DS}$ = 10 mA, the $I_S^{local} / I_S^{inj}$ value is increased to 47%, but more than the half (53%) is still wasted. To further increase $I_S^{local} / I_S^{inj}$, increase of $r_{NL}^{(S)}$ is needed.

**D. Spin detection efficiency of the D junction**

As the spin drift increases ($I_{DS}$ increases), $\lambda_{ch}^{d}$ and $r_{ch}^{u}$ increase. However, even though $r_{ch}^{u}$ becomes large enough ($r_{ch}^{u} \to \infty$) with a large spin drift velocity ($v_d \to \infty$), $r^{output}$ is limited by $r_{NL}^{(D)}$ ($r^{output} \to r_{NL}^{(D)}$). Indeed, as seen in Table 1, $r_{ch}^{u}$ is enhanced from 111 Ω to 467 Ω by the spin drift when $I_{DS}$ is increased from 2 to 10 mA, however, $r^{\text{output}}$ is almost unchanged (from 4.3 Ω to 4.5 Ω) due to the low $r_{\text{NL}}^{(\text{D})}$. This means that the most part of $I_{\text{S}}^{\text{trans}}$ diffuses toward the nonlocal region ($y > L_{\text{ch}}$) and the spin accumulation at $y = L_{\text{ch}}$ is reduced, which results in the reduction of the MR ratio. Thus, increase of $r_{\text{NL}}^{(\text{D})}$ in needen to increase $r^{\text{output}}$. Note that when $r^{\text{output}}$ exceeds $r_{\text{B}}^{(\text{D})}$, spin accumulation under the D electrode decreases due to the spin current that flows out from the $n^+$-Si region to the D electrode. Thus, the condition $r^{\text{output}} \sim r_{\text{B}}^{(\text{D})}$ gives the maximum MR ratio (see Section S6 in S.M. for the detail [25]).

**E. Design of spin MOSFET with a higher magnetoresistance ratio**

From the above-mentioned discussions, we obtained the following guideline to improve the performance of spin MOSFETs. We must realize a) – f),

a) High $P_{\text{S}}$ to increase the spin injection/detection polarization.

b) Matching the conductivity of the tunnel barrier to satisfy $r_{\text{B}}^{(\text{S})} \sim r^{\text{input}}$ and $r_{\text{B}}^{(\text{D})} \sim r^{\text{output}}$.

c) Long $\lambda_{\text{ch}}^{\text{d}}$ ($>>L_{\text{ch}}$) to reduce the spin-flip during the spin transport.

d) Low $r_{\text{ch}}^{\text{d}}$ to increase the local spin current ratio $I_{\text{S}}^{\text{local}} / I_{\text{S}}^{\text{inj}}$ and to reduce $r_{\text{B}}^{(\text{S})}$.

e) High $r_{\text{ch}}^{\text{u}}$ to enhance the spin accumulation at the D junction.

f) High $r_{\text{NL}}^{(\text{S})}$ and $r_{\text{NL}}^{(\text{D})}$ to reduce the spin diffusion towards the nonlocal regions.

It should be emphasized here that c), d), and e) have been realized in our spin MOSFET by the spin drift, while f) is not realized by the $n^+$-Si region formed at the S and D junctions.

We have introduced the $n^+$-Si regions into our spin MOSFET to enhance the MR ratio by reducing the parasitic resistance in the S/D junctions. However, it was found that these $n^+$-Si regions act as *sinks* of spins even though their thickness $t_n$ = 5 nm is far shorter than the spin diffusion length $\lambda_n$ = 0.46 μm. This is because electron spins diffuse along the $y$ direction in the $n^+$-Si regions with the lateral length ($L_S$ = 0.7 and $L_D$ = 2.0 μm) larger than $\lambda_n$ and spins are flipped significantly during the diffusion in the $n^+$-Si regions. One possible method to reduce such loss of spins is to use a lower doping concentration of the $n^+$-Si region. However, it results in lowering the MR ratio by the increase of the total channel resistance with higher Schottky barrier resistances formed in the S/D junctions. Thus, a promising method to reduce the loss of spins in the $n^+$-Si region and to increase the MR ratio simultaneously is the decrease of the length of the $n^+$-Si regions ($L_S$ and $L_D$) along the $y$ direction. When $L_S$ = 0.1 μm, as shown the bottom line in Table. 1, $r_{NL}^{(S)}$ increases to 14.7 Ω and the ratio $I_S^{local}/I_S^{inj}$ is enhanced by 1.6 times (from 47% to 77%). Note that even when $r_{NL}^{(S)}$ increases, $r^{input}$ keeps a lower value ( = 3.4 Ω) thanks to the low $r_{ch}^{d}$, so the conductivity matching condition does not change. On the other hand, when $L_D$ = 0.1 μm, $r^{output}$ is enhanced by 3.2 times (from 4.5 Ω to 14.3 Ω) as shown in Table. 1, and thus the spin-valve signal is enhanced by 3.2 times. From our calculation, when the lateral lengths of the both $n^+$-Si regions are 0.1 μm ($L_S$ = $L_D$ = 0.1 μm), spin-valve signal becomes 5.2 (= 1.6 × 3.2) times larger than our maximum value (~0.02%) with $L_{ch}$ = 0.4 μm, and the MR ratio will be increased to ~0.1%.

## VI. Conclusion

We fabricated Si-based spin MOSFETs with a lightly-$p$ Si channel and Fe/Mg/MgO/$n^+$-Si source/drain junctions and investigated the spin transport phenomena in the Si 2D inversion channel experimentally and theoretically. The $n^+$-Si regions were introduced to improve both the transistor characteristics and MR ratio by reducing the Schottky barrier resistance at the S/D junctions. We demonstrated the basic spin MOSFET operations at room temperature; transistor characteristics with a high on-off ratio (~$10^6$) and clear spin-valve signals. The maximum MR ratio was 0.02% in the spin MOSFET with $L_{ch}$ = 0.4 μm, which was enhanced by ~6 times than that in our previous paper [8].

To understand and analyze the experimental results in our spin MOSFETs, we developed the spin transport model by taking into account both the $n^+$-Si regions and the spin drift in the channel. The 2T Hanle signals were clearly observed in the spin MOSFET with $L_{ch}$ = 10 μm, which were perfectly reproduced by our analytical formula. From the fitting results, we obtained $P_S$ = 5.6 – 7.3%, $D_e^{ch}$ = 10.8 – 11.3 cm$^2$/s, $\tau_S^{ch}$ = 0.77 – 0.82 ns, and $\lambda_{ch}$ = 0.91 – 0.96 μm for $V_{GS}$ = 40 – 80 V. Since $\tau^{ch}$ = 32 – 34 fs, the momentum/spin lifetime was estimated to be $\tau^{ch} / \tau_S^{ch}$ ~ 1/25000. This value is somehow smaller than that in our previous spin MOSFET ($\tau^{ch} / \tau_S^{ch}$ ~ 1/14000) [8] having a $n$-type Si channel with a phosphorus donor concentration of $10^{17}$ cm$^{-3}$. This suggests that the strength of the spin-orbit coupling (SOC) is reduced by the lack of ionized phosphorus donors in the channel in the present spin MOSFETs.

Furthermore, we originally introduced the effective spin resistances in the spin MOSFET; $r_{NL}^{(S)}$, $r_{NL}^{(D)}$, $r_{ch}^{u}$, $r_{ch}^{d}$, $r^{input}$, and $r^{output}$, which allow us to understand the spin transport physics and to improve the performance of the spin MOSFETs. By using these spin resistances, the spin injection and detection efficiencies can be easily analyzed without any complex calculation. Through our experiments and analyses, we have obtained the guideline a) – f) to improve the performance of spin MOSFETs, as described in section V.E.

It was found that, in our spin MOSFETs, the injected spins are substantially flipped in the $n^+$-Si regions of the S/D junctions even though their thickness is far shorter (~5 nm) than the spin diffusion length (~ 0.5 μm). To solve this problem, we showed that the device structure should be modified to have large $r_{NL}^{(S)}$ and $r_{NL}^{(D)}$. One method is to reduce the lateral lengths of the both $n^+$-Si regions $L_S$ and $L_D$. When $L_S = L_D = 0.1$ μm, the MR ratio is expected to be increased to 0.1% in a spin MOSFET with $L_{ch} = 0.4$ μm.

There have been studies on spin MOSFETs for recent years, but none of them has focused on the device structure. We found that the optimization of the device structure including the $n^+$-Si regions of the S/D junctions is very effective to improve the MR ratio of spin MOSFETs. In this study, we clarified the design guideline for spin MOSFETs utilizing electron spin transport, which is different from that for the ordinary MOSFETs utilizing electron charge transport. Our findings enable researchers to analyze the spin current distributions in spin MOSFETs and show the guideline of the device design to realize high MR ratios.

**Acknowledgements**

We would like to express special thanks to M. Ichihara for his help in the device fabrication. This work was partially supported by Grants-in-Aid for Scientific

Research (No. 16H02095 and 18H05345), CREST of JST (No. JPMJCR1777), Yazaki Science and Technology Foundation, and Spintronics Research Network of Japan.

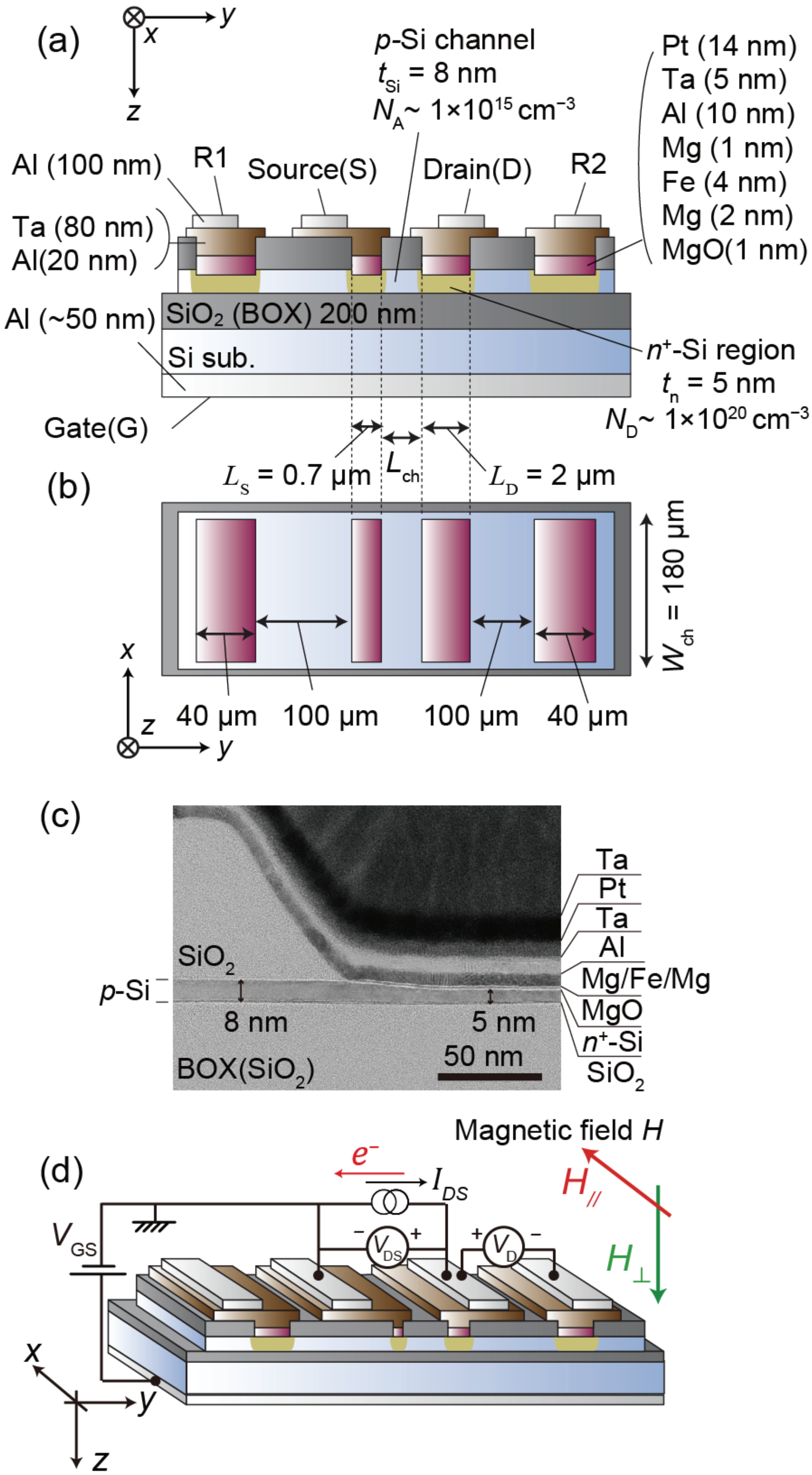

Figure 1 (a) Side view and (b) top view of a spin MOSFET structure having a 8-nm-thick $p$-Si channel and Fe(4 nm)/Mg(2 nm)/MgO(1 nm)/$n^+$-Si junctions prepared on a silicon-on-insulator (SOI) substrate, in which the accepter doping concentration $N_A$ of the $p$-Si channel is ~$1\times10^{15}$ cm$^{-3}$, the phosphorus donor doping concentration $N_D$ of the $n^+$-Si region (yellow area) is ~$1\times10^{20}$ cm$^{-3}$, and the thickness of the buried oxide (BOX) $SiO_2$ layer is 200 nm. The Cartesian coordinate is defined as follows: $x$ and $y$ are parallel to the longitudinal and transverse directions of the Source (S)/Drain (D) electrode, respectively, and $z$ is normal to the substrate plane. The channel length along the $y$ direction is $L_{ch}$ = 0.4 or 10 μm, the channel width along the $x$ direction is $W_{ch}$ = 180 μm, and the short side lengths along the $y$ direction of the S and D electrodes are $L_S$ = 0.7 μm and $L_D$ = 2.0 μm, respectively. The R1 and R2 electrodes are located at ~100 μm away from the S and D electrodes, respectively, and their short-side lengths along the $y$ direction is 40 μm. (c) Cross-sectional transmission electron microscopy (TEM) image around the left-hand side edge of the D electrode, where the electron beam incidence is along the $x$ axis ([110] axis of Si). (d) Two-terminal (2T) measurement setup of the spin MOSFET, where the voltage between the S and D (the D and R2) electrodes are measured simultaneously by the voltage meters $V_{DS}$ ($V_D$) while a constant current $I_{DS}$ is driven from the D to S electrodes and a constant positive gate-source voltage $V_{GS}$ is applied to the back side with respect to the grounded S electrode. In spin-valve measurements, an external magnetic field is applied along the plane ($H = H_{//}$, along the $x$ direction). In 2T Hanle measurements, an external magnetic field is applied perpendicular to the plane ($H = H_{\perp}$, along the $z$ direction).

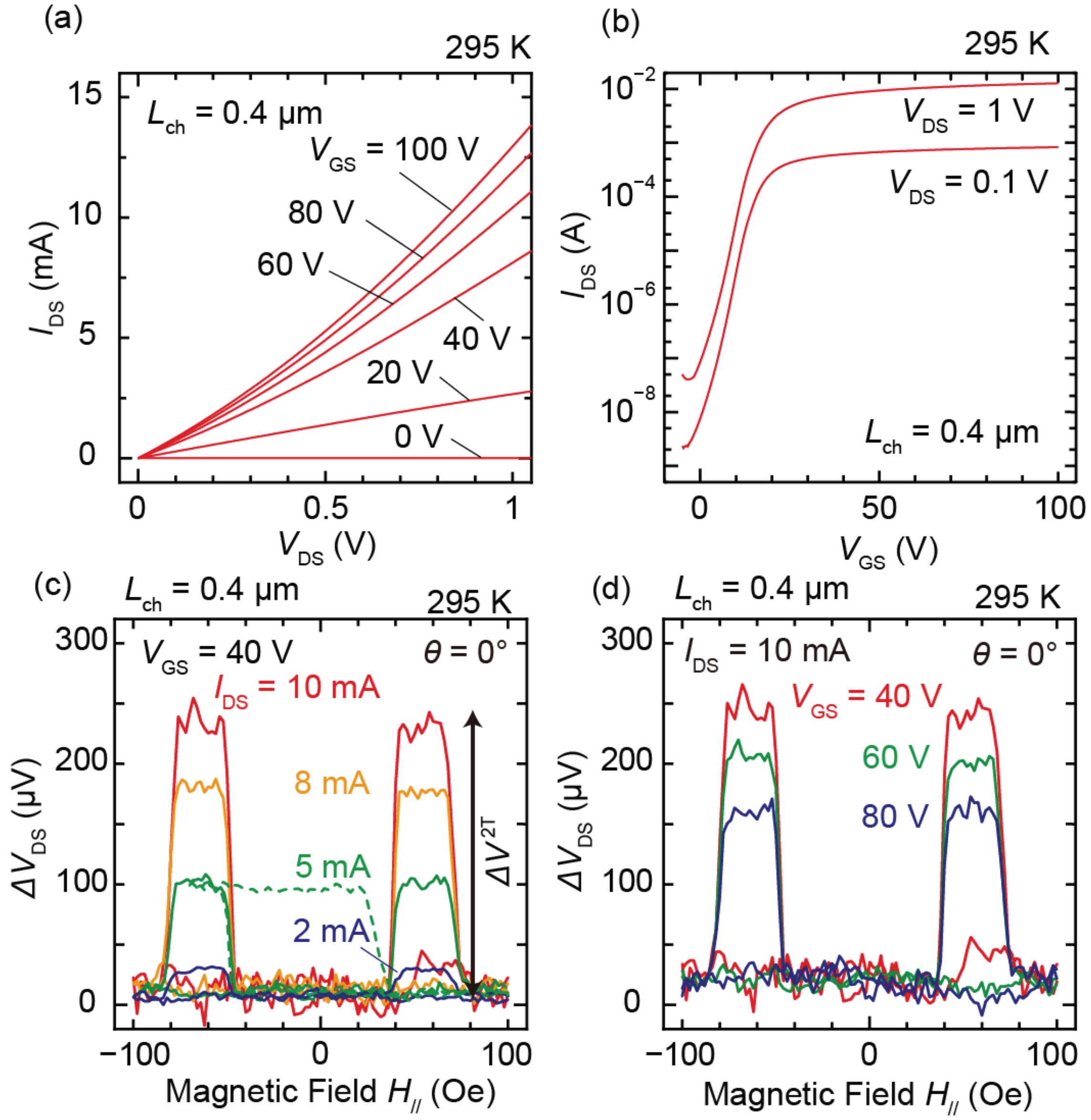

(a)
295 K
$L_{ch}$ = 0.4 μm
$V_{GS}$ = 100 V
80 V
60 V
40 V
20 V
0 V
$I_{DS}$ (mA)
$V_{DS}$ (V)
(b)
295 K
$V_{DS}$ = 1 V
$V_{DS}$ = 0.1 V
$L_{ch}$ = 0.4 μm
$I_{DS}$ (A)
$V_{GS}$ (V)
(c)
$L_{ch}$ = 0.4 μm
295 K
$V_{GS}$ = 40 V
θ = 0°
$I_{DS}$ = 10 mA
8 mA
5 mA
2 mA
$\Delta V^{2T}$
$\Delta V_{DS}$ (μV)
Magnetic Field $H_{//}$ (Oe)
(d)
$L_{ch}$ = 0.4 μm
295 K
$I_{DS}$ = 10 mA
θ = 0°
$V_{GS}$ = 40 V
60 V
80 V
$\Delta V_{DS}$ (μV)
Magnetic Field $H_{//}$ (Oe)

Figure 2 (a) $I_{DS}$ - $V_{DS}$ characteristics measured at 295 K for a spin MOSFET with $L_{ch}$ = 0.4 μm, where $V_{GS}$ was varied from 0 to 100 V in the step of 20 V. (b) $I_{DS}$ - $V_{GS}$ characteristics measured at 295 K for the same device, where $V_{DS}$ = 0.1 V and 1 V. (c) Voltage change $\Delta V_{DS}$ measured at 295 K with various $I_{DS}$ and $V_{GS}$ = 40 V for the same device, while an in-plane magnetic field ($H = H_{//}$) is swept. The blue, green, orange, and red solid curves are the major loops measured with $I_{DS}$ = 2, 5, 8, and 10 mA, respectively, and the green dashed curve is the minor loop measured with $I_{DS}$ = 5 mA. The amplitude of the spin-valve signal $\Delta V^{2T}$ is defined as the voltage difference between antiparallel and parallel magnetization configurations. (d) Voltage change $\Delta V_{DS}$ measured at 295 K with various $V_{GS}$ and $I_{DS}$ = 10 mA for the same device. The red, green, and blue solid curves are major loops measured with $V_{GS}$ = 40, 60, and 80 V, respectively.

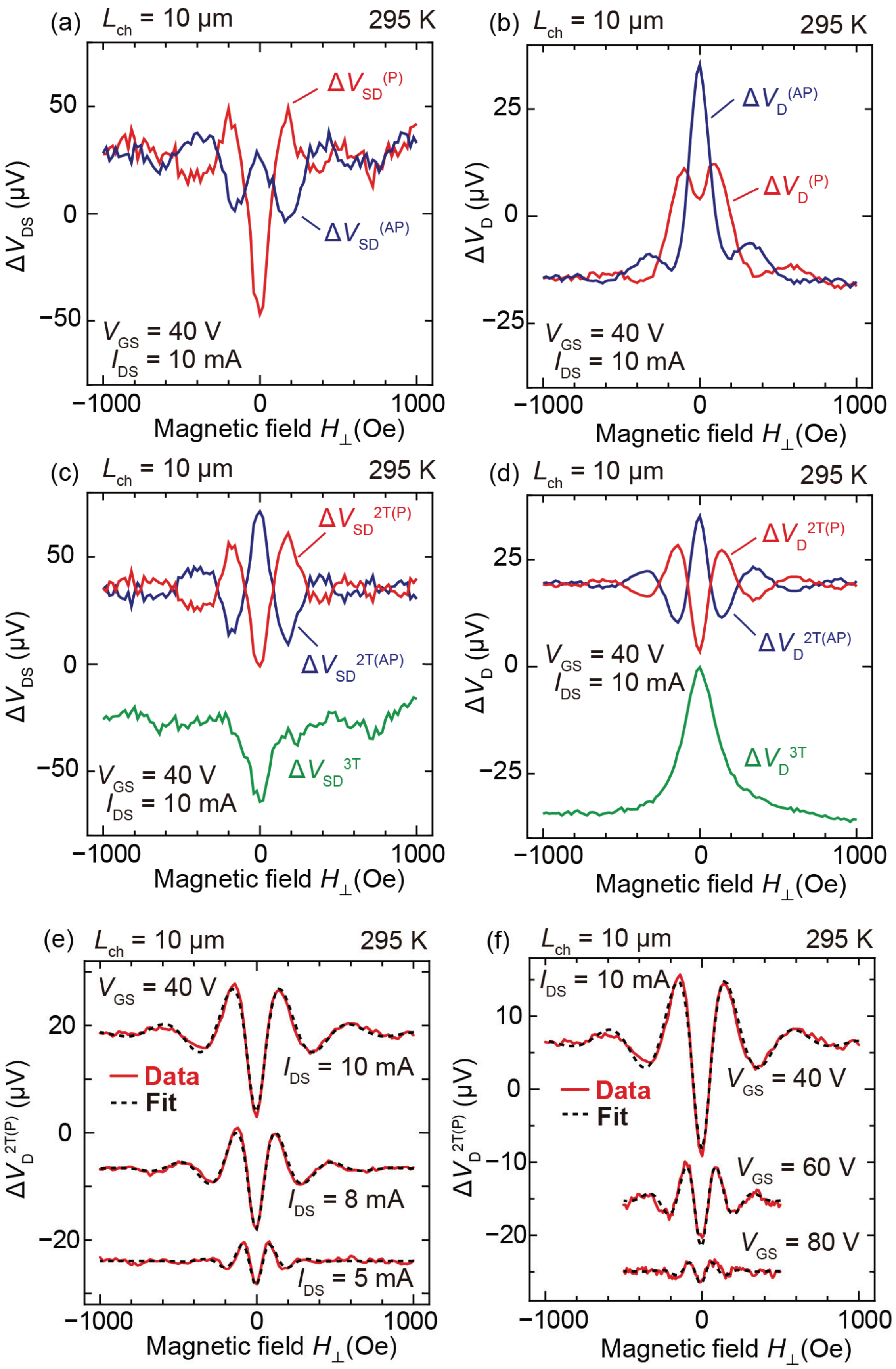

(a) $L_{ch}$ = 10 μm
295 K
$\Delta V_{SD}^{(P)}$
$\Delta V_{SD}^{(AP)}$
$V_{GS}$ = 40 V
$I_{DS}$ = 10 mA
$\Delta V_{DS}$ (μV)
50
0
−50
−1000
0
1000
Magnetic field $H_\perp$(Oe)
(b) $L_{ch}$ = 10 μm
295 K
$\Delta V_D^{(AP)}$
$\Delta V_D^{(P)}$
$V_{GS}$ = 40 V
$I_{DS}$ = 10 mA
$\Delta V_D$ (μV)
25
0
−25
−1000
0
1000
Magnetic field $H_\perp$(Oe)
(c) $L_{ch}$ = 10 μm
295 K
$\Delta V_{SD}^{2T(P)}$
$\Delta V_{SD}^{2T(AP)}$
$\Delta V_{SD}^{3T}$
$V_{GS}$ = 40 V
$I_{DS}$ = 10 mA
$\Delta V_{DS}$ (μV)
50
0
−50
−1000
0
1000
Magnetic field $H_\perp$(Oe)
(d) $L_{ch}$ = 10 μm
295 K
$\Delta V_D^{2T(P)}$
$\Delta V_D^{2T(AP)}$
$\Delta V_D^{3T}$
$V_{GS}$ = 40 V
$I_{DS}$ = 10 mA
$\Delta V_D$ (μV)
25
0
−25
−1000
0
1000
Magnetic field $H_\perp$(Oe)
(e) $L_{ch}$ = 10 μm
295 K
$V_{GS}$ = 40 V
Data
Fit
$I_{DS}$ = 10 mA
$I_{DS}$ = 8 mA
$I_{DS}$ = 5 mA
$\Delta V_D^{2T(P)}$ (μV)
20
0
−20
−1000
0
1000
Magnetic field $H_\perp$(Oe)
(f) $L_{ch}$ = 10 μm
295 K
$I_{DS}$ = 10 mA
Data
Fit
$V_{GS}$ = 40 V
$V_{GS}$ = 60 V
$V_{GS}$ = 80 V
$\Delta V_D^{2T(P)}$ (μV)
10
0
−10
−20
−1000
0
1000
Magnetic field $H_\perp$(Oe)

Figure 3 (a)(b) Voltage change (a) $\Delta V_{DS}$ and (b) $\Delta V_D$ measured at 295 K with $I_{DS}$ = 10 mA and $V_{GS}$ = 40 V for a spin MOSFET with $L_{ch}$ = 10 μm, while a perpendicular-to-plane magnetic field ($H = H_{\perp}$) is swept. Red and blue curves are row data in the parallel (P) and antiparallel (AP) magnetization configuration, respectively. (c)(d) Extracted 2T and 3T Hanle signals, which were obtained using the data in (a)(b). Red and blue curves are extracted 2T Hanle signals in the parallel and antiparallel magnetization configuration using Eqs. (1b) and (1c), respectively, and the bottom green curves are extracted 3T Hanle signals using Eq. (1a). (e)(f) Extracted 2T Hanle signals $\Delta V_D^{2T(P)}$ (red solid curves) and the fitting results of Eq. (4a) (black dashed curve). (e) Bias current dependence measured with $I_{DS}$ = 10, 8, and 5 mA and the same $V_{GS}$ = 40 V. (f) Gate voltage dependence measured with $V_{GS}$ = 40, 60, and 80 V and the same $I_{DS}$ = 10 mA. Each graph is vertically shifted for clear vision.

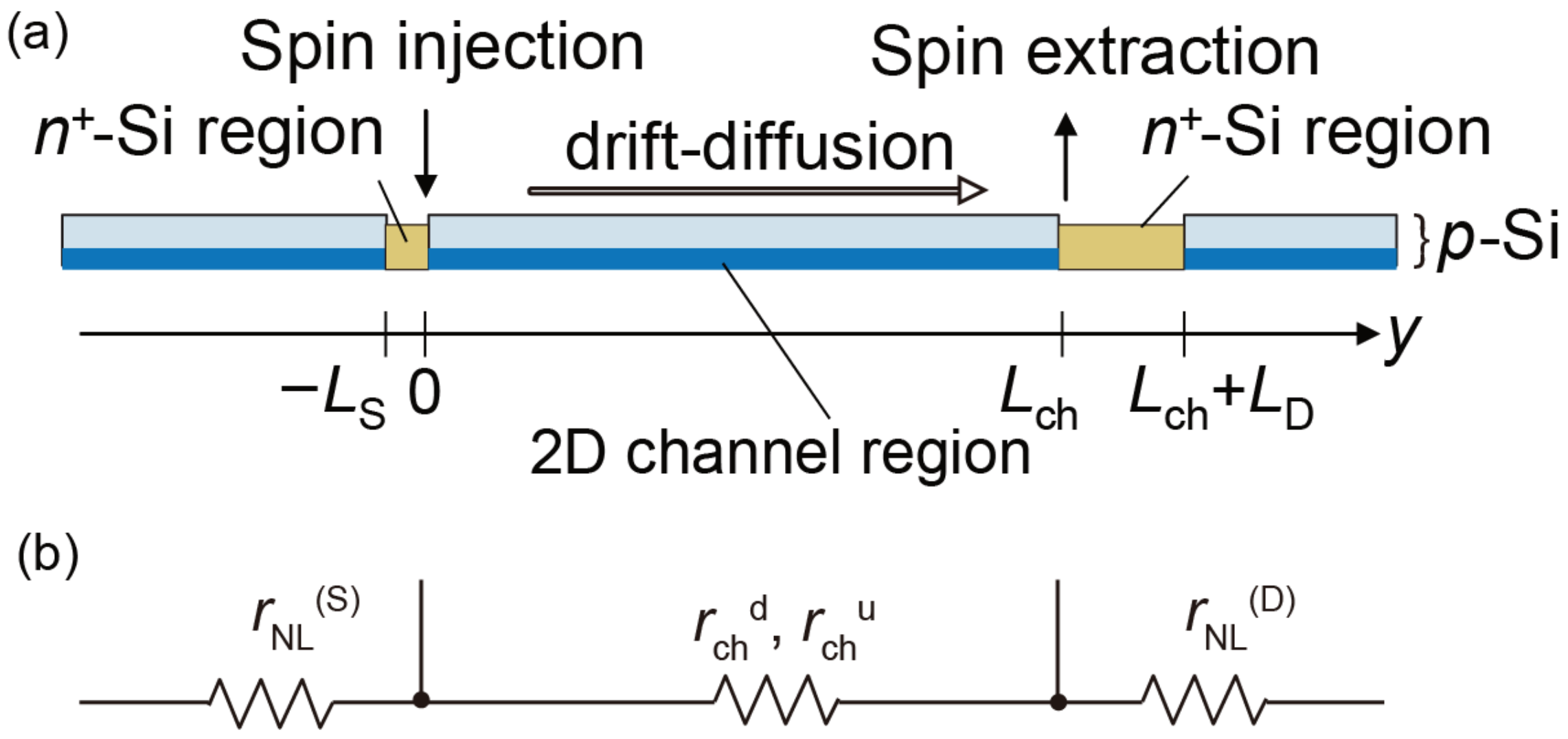

Figure 4 (a) Schematic illustration of our model for the spin transport through the channel in our spin MOSFETs, where light blue, dark blue, and yellow regions represent a *p*-Si, 2D inversion channel, and $n^+$-Si regions, respectively. The *y*-axis is defined along the spin transport direction in the device plane and its origin is located at the right-hand edge of the S electrode. Spins are injected and extracted at the right-hand side of the S electrode ($y$ = 0) and the left-hand side of the D electrode ($y = L_{ch}$), respectively, and the spin drift occurs only between S and D ($0 < y < L_{ch}$). (b) Equivalent circuit expressed by the spin resistances, where $r_{ch}^{d}$, $r_{ch}^{u}$, $r_{NL}^{(S)}$ and $r_{NL}^{(D)}$ are defined in Eqs. (5a)–(5d). $r_{ch}^{d}$ and $r_{ch}^{u}$ are for spins that flow positive and negative direction, respectively.

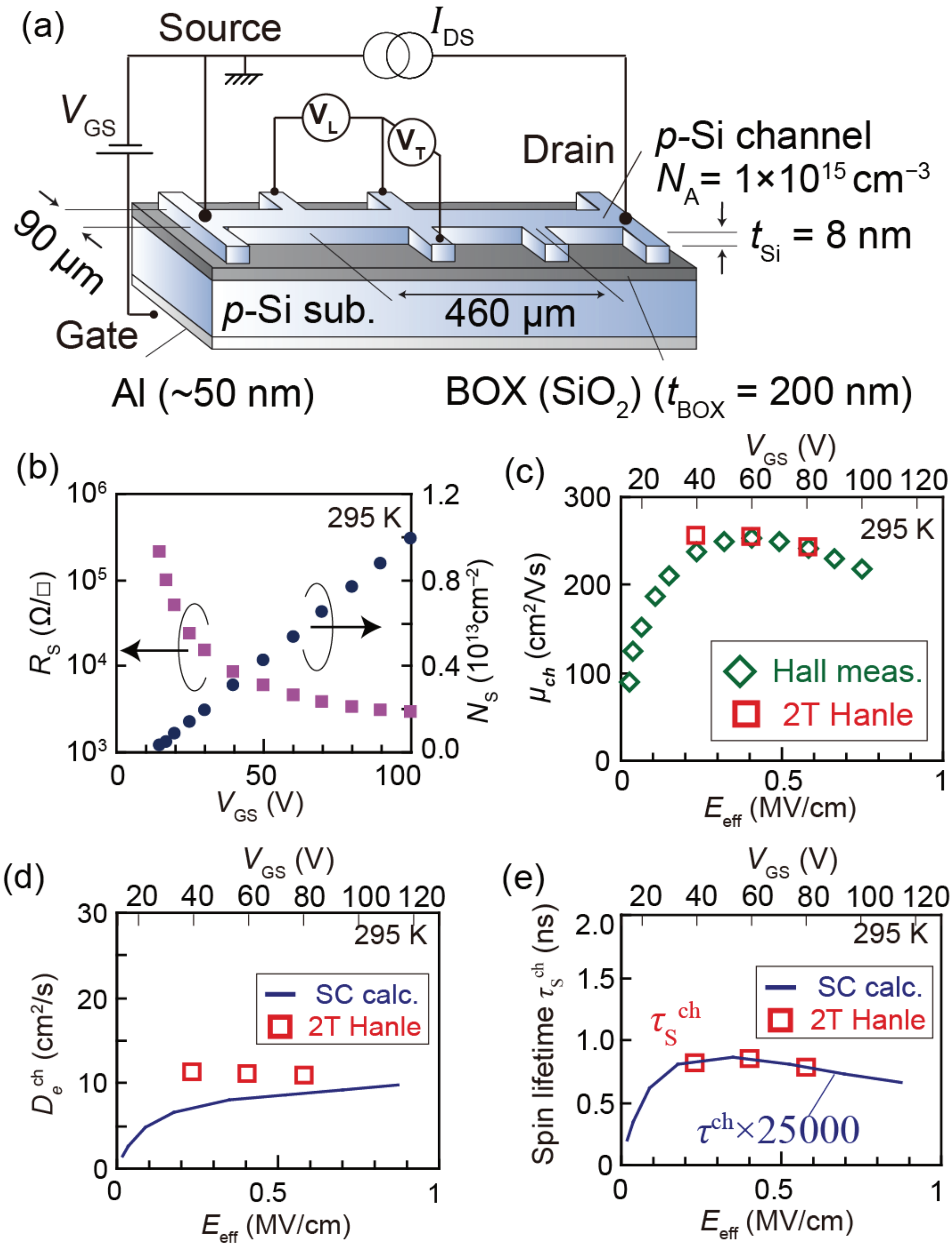

Figure 5 (a) Schematic illustration of a Hall-bar-type MOSFET (type-I) prepared on a SOI substrate, where the channel thickness is $t_{\mathrm{Si}}$ = 8 nm, the accepter doping concentration $N_{\mathrm{A}}$ is $1\times10^{15}$ cm$^{-3}$, and the channel length and width are 460 μm and 90 μm, respectively. The measurement setup is also shown, where a constant drain-source current $I_{\mathrm{DS}}$ and a constant gate-source voltage $V_{\mathrm{GS}}$ were applied and the longitudinal voltage $V_{\mathrm{L}}$ and transverse voltages $V_{\mathrm{T}}$ are measured while a sweeping magnetic field was applied perpendicular to the substrate plane. (b) Channel sheet resistance $R_{\mathrm{S}}$ (left axis) and sheet electron density $N_{\mathrm{S}}$ (right axis) in the Si 2D inversion channel plotted as a function of $V_{\mathrm{GS}}$ estimated from Hall measurements at 295 K. (c) Effective electron mobility $\mu_{\mathrm{ch}}$ in the inversion channel plotted as a function of effective gate electric field $E_{\mathrm{eff}} = q / \varepsilon_{\mathrm{Si}}(N_{\mathrm{S}} / 2 + N_{\mathrm{A}} t_{\mathrm{Si}})$, where green open diamonds are those estimated from Hall measurements and red open squares are those estimated form the 2T Hanle signals in Figs. 3(d). (d) Electron diffusion coefficient $D_e^{\mathrm{ch}}$ plotted as a function of $E_{\mathrm{eff}}$, where a blue solid curve was estimated from a self-consistent calculation (see S3 in S.M.[25]) and red open squares were estimated form the 2T Hanle signals. (e) Spin lifetime $\tau_{\mathrm{S}}^{\mathrm{ch}}$ in the 2D inversion channel plotted as a function of effective gate electric field $E_{\mathrm{eff}}$, where red open squares were estimated from the 2T Hanle signals. The blue curve is the momentum lifetime $\tau^{\mathrm{ch}}$ multiplied by 25000, which was estimated from the self-consistent calculation (see S3 in S.M. [25]).

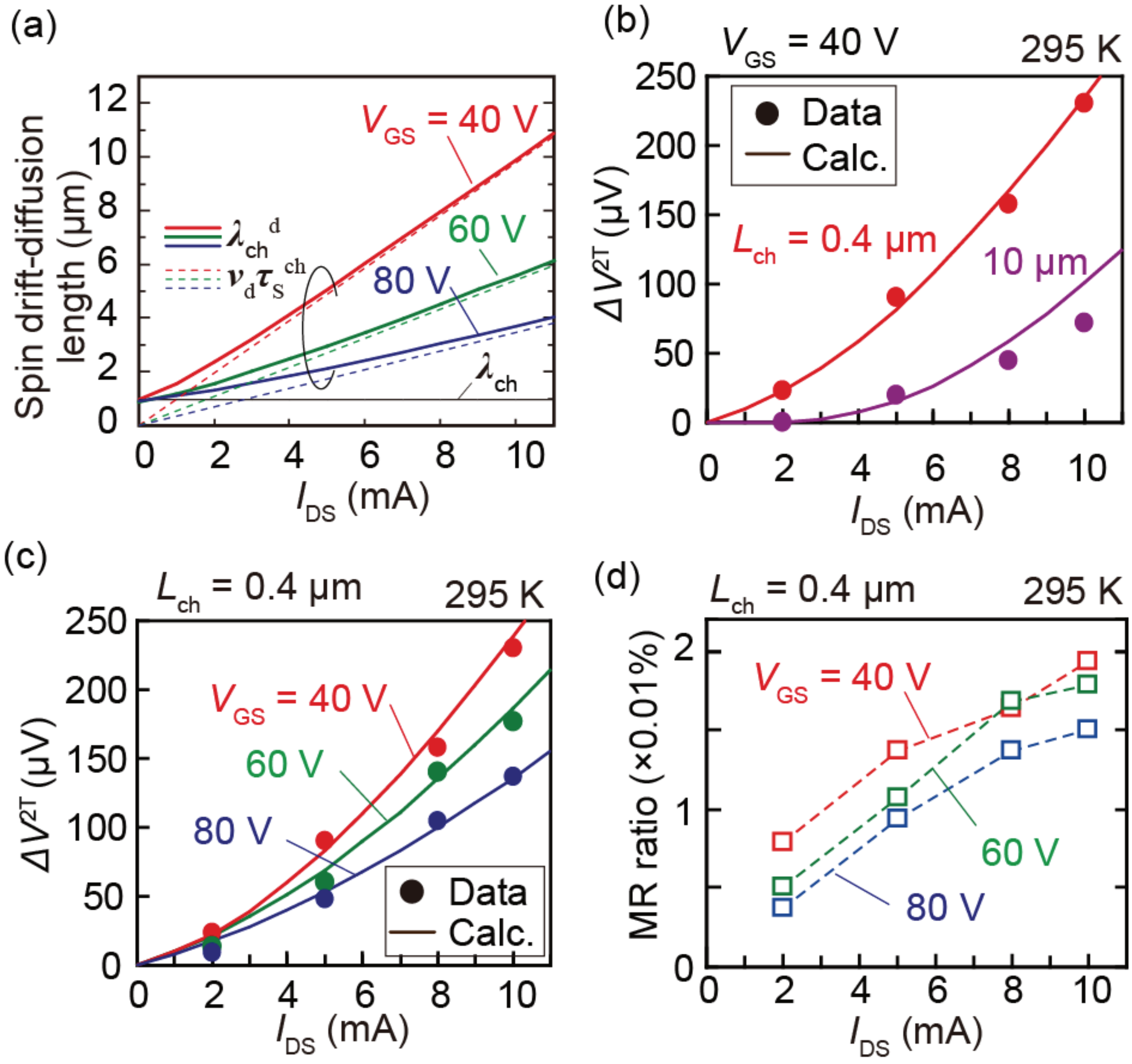

(a)
Spin drift-diffusion length (μm)
V_GS = 40 V
60 V
80 V
λ_ch^d
v_d τ_S^ch
λ_ch
I_DS (mA)
(b)
V_GS = 40 V
295 K
Data
Calc.
L_ch = 0.4 μm
10 μm
ΔV^2T (μV)
I_DS (mA)
(c)
L_ch = 0.4 μm
295 K
V_GS = 40 V
60 V
80 V
Data
Calc.
ΔV^2T (μV)
I_DS (mA)
(d)
L_ch = 0.4 μm
295 K
V_GS = 40 V
60 V
80 V
MR ratio (×0.01%)
I_DS (mA)

Figure 6 (a) Spin drift-diffusion lengths as a function of $I_{DS}$ calculated by Eq. (4e), where red, green, and blue solid curves are down-stream spin drift-diffusion length ($\lambda_{ch}^{d}$) at $V_{GS}$ = 40, 60, and 80 V, respectively, red, green, and blue dashed lines are spin drift length ($v_d \tau_S^{ch}$) at $V_{GS}$ = 40, 60, and 80 V, respectively, and a black solid line is the intrinsic spin diffusion length $\lambda_{ch}$ in the channel. (b) $\Delta V^{2T}$ as a function of $I_{DS}$ when signals were measured for the spin MOSFETs with $L_{ch}$ = 0.4 (Fig. 2(c)) and 10 μm at $V_{GS}$ = 40 V. Red and purple filled circles are experimental data for $L_{ch}$ = 0.4 and 10 μm, respectively, and red and purple solid curves are theoretical calculations by Eq. (6) with $P_S$ = 7.3%. (c) $\Delta V^{2T}$ as a function of $I_{DS}$ when signals were measured for the spin MOSFETs with $L_{ch}$ = 0.4 μm at various $I_{DS}$ (= 2, 5, 8, and 10 mA) and $V_{GS}$ (= 40, 60, and 80 V). Red, green, and blue filled circles are experimental data (Fig. 2(d)) measured at $V_{GS}$ = 40, 60, and 80 V, respectively, and red, green, and blue solid curves are theoretical calculations by Eq. (6) with $P_S$ = 7.3%, 6.5 %, and 5.6%, respectively. (d) MR ratio obtained in the spin MOSFETs with $L_{ch}$ = 0.4 μm, which were obtained using the data in (c). Red, green, and blue open squares are values for $V_{GS}$ = 40, 60, and 80 V, respectively, and dashed lines are guides for eyes.

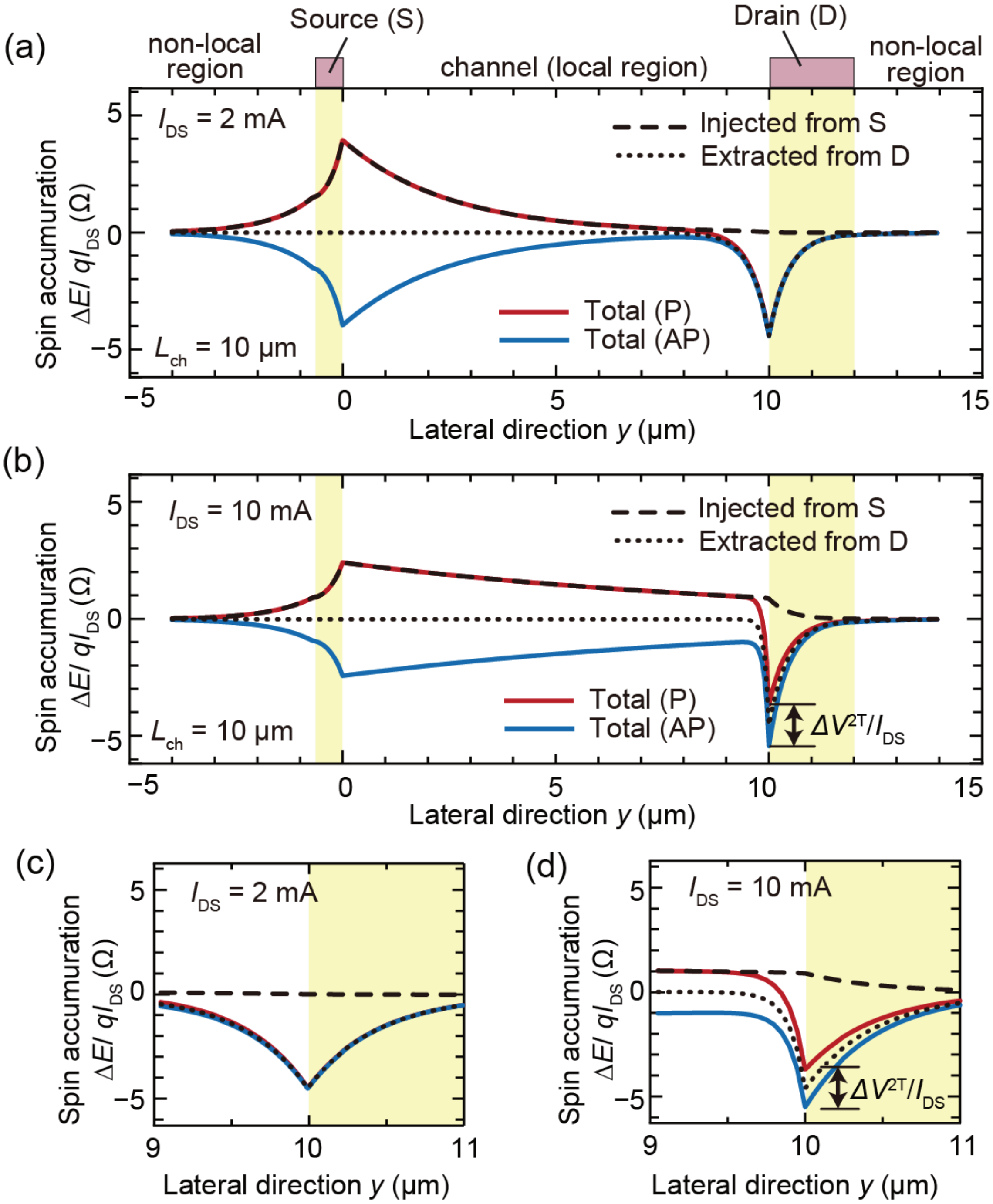
(a)
Source (S)
Drain (D)
non-local region
channel (local region)
non-local region
$I_{DS}$ = 2 mA
Injected from S
Extracted from D
Total (P)
Total (AP)
$L_{ch}$ = 10 µm
Spin accumuration
ΔE/ qI$_{DS}$ (Ω)
Lateral direction y (µm)
(b)
$I_{DS}$ = 10 mA
Injected from S
Extracted from D
Total (P)
Total (AP)
$ΔV^{2T}/I_{DS}$
$L_{ch}$ = 10 µm
Spin accumuration
ΔE/ qI$_{DS}$ (Ω)
Lateral direction y (µm)
(c)
$I_{DS}$ = 2 mA
Spin accumuration
ΔE/ qI$_{DS}$ (Ω)
Lateral direction y (µm)
(d)
$I_{DS}$ = 10 mA
$ΔV^{2T}/I_{DS}$
Spin accumuration
ΔE/ qI$_{DS}$ (Ω)
Lateral direction y (µm)

Figure 7 (a)(b) Distribution of the normalized spin accumulation $\Delta E / qI_{DS}$ in the channel with $L_{ch}$ = 10 μm when (a) $I_{DS}$ = 2 mA and (b) 10 mA, which were calculated by solving Eq. (2) with $P_S$ = 1 (see S2 in S.M. [25]). Red and blue solid curves represent the total spin accumulations in the parallel (P) and an antiparallel (AP) configurations, respectively, a black dashed curve is the components of the spins injected from the S electrode, and a black dotted curve is the components of the spins extracted from the D electrode. (c)(d) Close-up views at around $y$ = 10 μm (near the D electrode) of (a)(b). Black double-headed arrows in (b) and (d) denote the spin-valve signal observed by the D electrode.

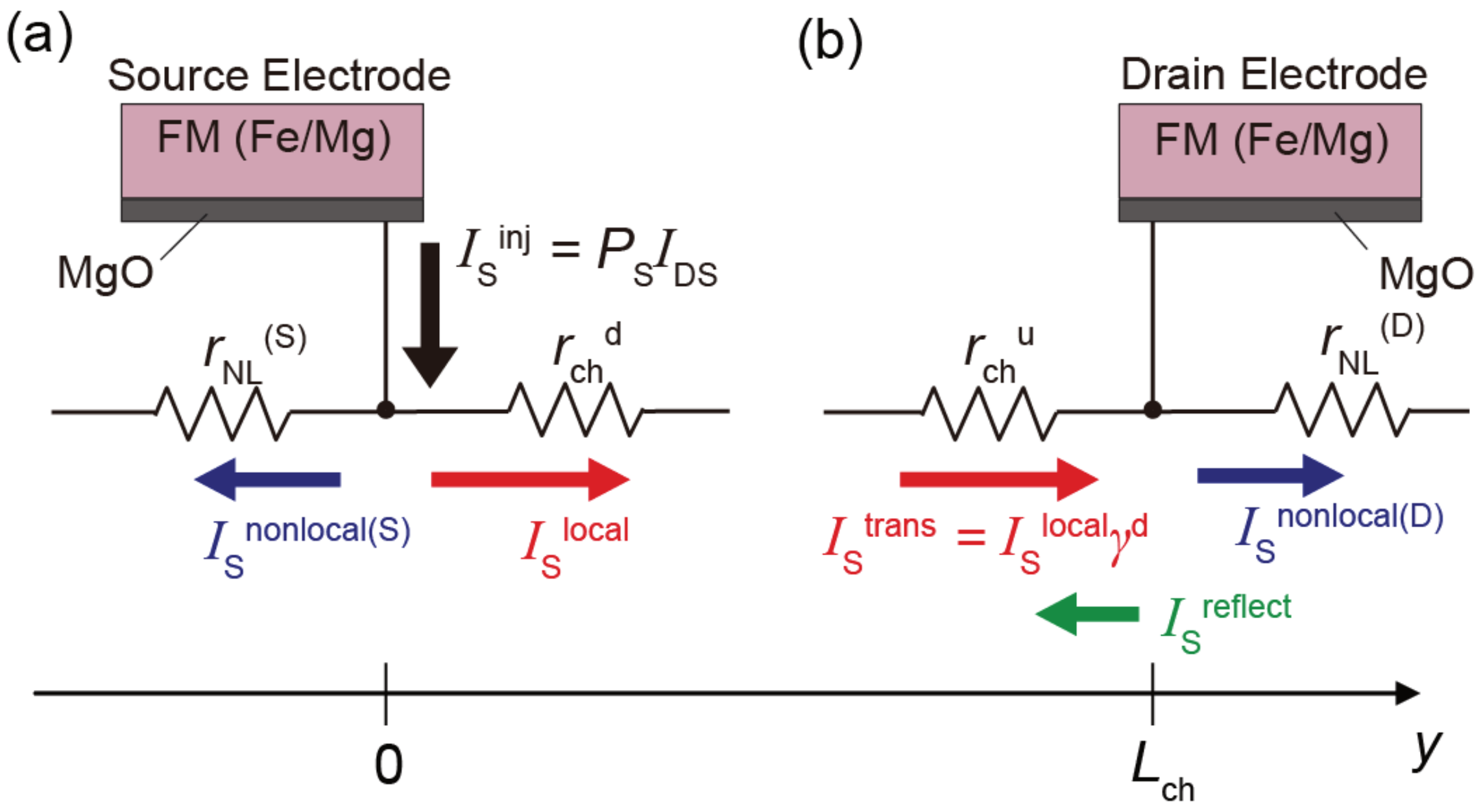

Figure 8 (a)(b) Equivalent circuit consisting of the spin resistances around the S and D electrode, respectively, in which $r_{ch}^{d}$, $r_{ch}^{u}$, $r_{NL}^{(S)}$ and $r_{NL}^{(D)}$ are defined in Eqs. (5a)–(5d). $I_S^{inj} = P_S I_{DS}$ is the spin current injected from the S electrode, $I_S^{local}$ is the spin current that flows into the local region ( $y > 0$ ), and $I_S^{nonlocal(S)}$ is the spin current that flows into the nonlocal region ( $y < 0$ ), $I_S^{trans} = I_S^{local}\gamma^d$ is the spin current transported from the S electrode through the channel, $\gamma^d$ is the decay factor defined in Eq. (4c), $I_S^{nonlocal(D)}$ is the spin current that flows into the nonlocal region ( $y > L_{ch}$ ), and $I_S^{reflect}$ is the spin current that flows back to the local region ( $y < L_{ch}$ ).

Table 1  Spin resistances in our spin MOSFET at $V_{GS}$ = 40 V for different $I_{DS}$ = 2 mA and 10 mA calculated using Eqs. (5a)−(5d), (7), and (8). The ratio $I_{\mathrm{S}}^{\mathrm{local}} / I_{\mathrm{S}}^{\mathrm{inj}} = r^{\mathrm{input}} / r_{\mathrm{ch}}^{\mathrm{d}}$ is also listed in the last column. Bottom row values are calculated by assuming $L_{\mathrm{S}} = L_{\mathrm{D}} = 0.1$ μm.

| $I_{DS}$ | $L_{\mathrm{S}} / L_{\mathrm{D}}$ | $r_{\mathrm{NL}}^{(\mathrm{S})}$ | $r_{\mathrm{NL}}^{(\mathrm{D})}$ | $r_{\mathrm{ch}}^{\mathrm{d}}$ | $r_{\mathrm{ch}}^{\mathrm{u}}$ | $r^{\mathrm{input}}$ | $r^{\mathrm{output}}$ | $I_{\mathrm{S}}^{\mathrm{local}} / I_{\mathrm{S}}^{\mathrm{inj}}$ |
|---|---|---|---|---|---|---|---|---|
| 2 mA | 0.7/2.0 μm | 4.9 Ω | 4.5 Ω | 18.5 Ω | 111 Ω | 3.9 Ω | 4.3 Ω | 21% |
| 10 mA | 0.7/2.0 μm | 4.9 Ω | 4.5 Ω | 4.4 Ω | 467 Ω | 2.3 Ω | 4.5 Ω | 47% |
| 10 mA | 0.1/0.1 μm | 14.7 Ω | 14.7 Ω | 4.4 Ω | 467 Ω | 3.4 Ω | 14.3 Ω | 77% |

**Supplemental Material**

# Spin transport in Si-based spin metal-oxide-semiconductor field-effect transistors: Spin drift effect in the inversion channel and spin relaxation in the $n^+$-Si source/drain regions

Shoichi Sato[1,2], Masaaki Tanaka[1,2], and Ryosho Nakane[1]

[1]*Department of Electrical Engineering and Information Systems, The University of Tokyo, 7-3-1 Hongo, Bunkyo-ku, Tokyo 113-8656, Japan*
[2]*Center for Spintronics Research Network (CSRN), The University of Tokyo, 7-3-1 Hongo, Bunkyo-ku, Tokyo 113-8656, Japan*

## S1. Current crowding at the D electrode

In the main manuscript, we measured the spin-valve signals and Hanle signals in the two-terminal (2T) measurement setup (Fig. 1(d)), where $V_{DS}$ is the voltage difference between the S and D electrodes and $V_D$ is that between the D and R2 electrodes. Theoretically, the signals $\Delta V_{DS}$ and $\Delta V_D$ obtained in $V_{DS}$ and $V_D$, respectively, should be the same [S1], and they were actually the same in some previous experimental studies [S2–S4]. However, we found that $\Delta V_D$ was almost half of $\Delta V_{DS}$ as shown in Figs. 3(c) and 3(d) in the main manuscript. The difference is probably caused by the current crowding at the left-hand edge of the D electrode, which was not taken into account in the basic theory. In this section, we estimate the current distribution near the D electrode using an equivalent circuit model and show that the difference between $\Delta V_D$ and $\Delta V_{SD}$ is attributed to the current crowding in our spin

MOSFETs.

Figure S1(a) shows a schematic picture around the D junction in the 2T measurement setup, in which the light blue, dark blue, and yellow regions are the lightly $p$-Si regon, Si inversion channel, and $n^+$-Si region, respectively, and the brown region is the MgO tunnel barrier. $R_{ch}$ is the channel resistance and $R_{ref}$ is the resistance between the D and R2 electrode. To simplify the discussion, the resistances of the S and R2 junctions are expressed by equivalent resistances $R_{Source}$ and $R_{R2}$, respectively. During the measurement, an electrical current flows from the D electrode to the S electrode through both the $n^+$-Si region and the channel region, whereas there is no electrical current in the nonlocal region ( $y > L_{ch} + L_D$ ).

When the resistance $R_N$ $(= \rho_n L_D / t_n W_{ch})$ of the $n^+$-Si region along the $y$ direction is much lower than the MgO tunnel resistance $R_{Drain}$, the electrical potential in the $n^+$-Si region becomes uniform, i.e., the tunnel current passing through the MgO tunnel barrier has a uniform distribution along the $y$ direction. In this case, the equivalent circuit around the D electrode can be expressed by Fig. S1(b). Here, $R_{Drain} = R_{D0} + \varDelta R_{Drain}$ is the MgO tunnel resistance, where $R_{D0}$ is that in the parallel magnetization configuration and $\varDelta R_{Drain}$ is the change in resistance due to the spin-valve effect. $V_{DS}$ and $V_D$ are expressed as follows:

$$V_{DS} = I_{DS}\left(R_{Source} + R_{ch} + R_{Drain}\right), \quad \text{(S1)}$$

$$V_D = I_{DS} R_{Drain}. \quad \text{(S2)}$$

Consequently, $\varDelta V_{DS}$ and $\varDelta V_D$ are;

$$\varDelta V_{DS} = I_{DS} \varDelta R_{Drain}, \quad \text{(S3)}$$

$$\varDelta V_D = I_{DS} \varDelta R_{Drain}. \quad \text{(S4)}$$

Therefore, the same amplitude of the spin-valve signals is obtained in $V_{DS}$ and $V_D$, as predicted by the basic theory [S1].

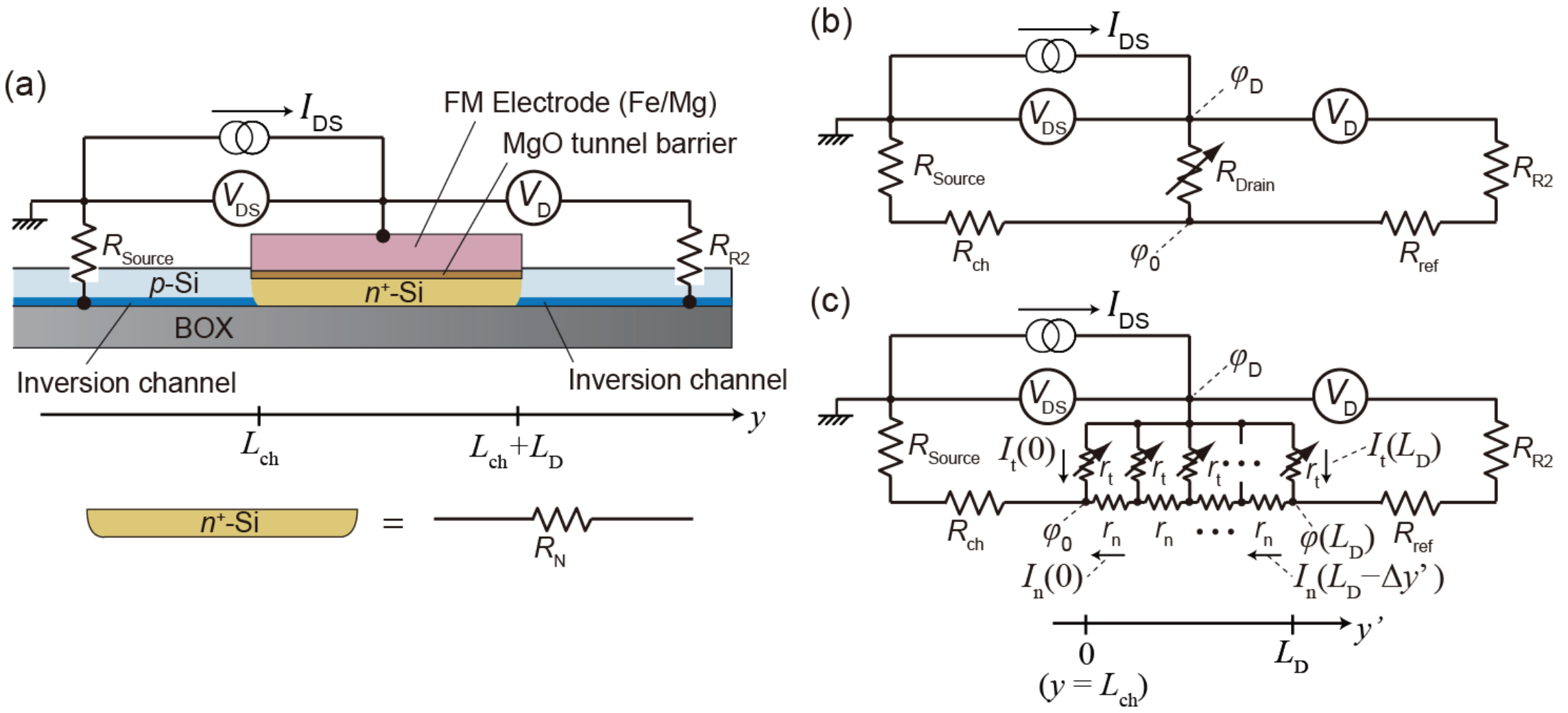

Figure S1 (a) Schematic picture of our spin MOSFET around the D junction in the 2T measurement setup. Light blue, dark blue, yellow, and brown regions represent the lightly $p$-Si region, inversion channel region, $n^+$-Si region, and the MgO tunnel barrier, respectively. $R_N$ is the resistance of the $n^+$-Si region along the $y$ direction. The resistances of the S and R2 junctions are represented by equivalent resistances $R_{Source}$ and $R_{R2}$, respectively. (b) Equivalent circuit of the D junction when $R_N << R_{Drain}$, where $R_{Drain}$ is the MgO tunnel resistance whose value is changed by the spin-valve effect. (c) Equivalent circuit of the D junction when $R_N$ is comparable to $R_{Drain}$. $r_t$ is the tunnel resistance of the infinitesimal length $\Delta y'$ and $r_n$ is the lateral resistance of the infinitesimal length $\Delta y'$ in the $n^+$-Si region. $\varphi(y')$ is the potential in the $n^+$-Si region, $\varphi_0 = \varphi(0)$, $\varphi_D$ is the potential of the FM electrode above the MgO tunnel barrier, $I_n(y')$ is the current in the $n^+$-Si region, and $I_t(y')$ is the tunnel current passing through the MgO tunnel barrier.

On the contrary, when $R_N$ is comparable to $R_{Drain}$, the voltage drop due to the lateral current flow in the $n^+$-Si region is not negligible and thus the tunnel current passing through the MgO tunnel barrier has a non-uniform distribution along the $y$ direction. In this case, the equivalent circuit of the D junction can be expressed by Fig. S1(c). We define $y'$ axis along the $y$ direction so that its origin is located at $y = L_{ch}$, namely, $y' = y - L_{\mathrm{ch}}$. In Fig. S1(c), $r_{\mathrm{t}} = R_{\mathrm{Drain}} L_{\mathrm{D}} / \Delta y'$ is the tunnel resistance of the infinitesimal length $\Delta y'$, $r_{\mathrm{n}} = \rho_{\mathrm{n}} \Delta y' / W_{\mathrm{ch}} t_{\mathrm{n}}$ is the lateral resistance of the infinitesimal length $\Delta y'$ in the $n^+$-Si region, $\varphi(y')$ is the potential in the $n^+$-Si region, $I_{\mathrm{n}}(y')$ is the current in the $n^+$-Si region, and $I_{\mathrm{t}}(y')$ is the tunnel current passing through the MgO tunnel barrier. Here, the vertical current flow (along the *z* direction) in the $n^+$-Si region was neglected since the thickness of the $n^+$-Si region (~5 nm) is much smaller than its length (~2 μm) along the *y* direction. Ohm's low and Kirchhoff's current law are expressed by;

$$I_{\mathrm{n}}(y') r_{\mathrm{n}} = \left( \varphi(y' + \Delta y') - \varphi(y') \right) / q , \tag{S5}$$

$$I_{\mathrm{t}}(y') r_{\mathrm{t}} = \left( \varphi_{\mathrm{D}} - \varphi(y') \right) / q , \tag{S6}$$

$$I_{\mathrm{n}}(y') = I_{\mathrm{n}}(y' + \Delta y') + I_{\mathrm{t}}(y' + \Delta y') \qquad (0 \le y' < L_{\mathrm{D}}), \tag{S7}$$

$$I_{\mathrm{DS}} = I_{\mathrm{n}}(0) + I_{\mathrm{t}}(0), \tag{S8}$$

$$I_{\mathrm{n}}(L_{\mathrm{D}} - \Delta y') = I_{\mathrm{t}}(L_{\mathrm{D}}), \tag{S9}$$

In the limit of $\Delta y' \to 0$, Eqs. (S5) – (S8) are solved as follows;

$$\varphi(y') = \varphi_{\mathrm{D}} - (\varphi_{\mathrm{D}} - \varphi_0) \frac{\cosh\left[ (L_{\mathrm{D}} - y') / l^* \right]}{\cosh\left( L_{\mathrm{D}} / l^* \right)}, \tag{S10}$$

$$\left| I_{\mathrm{n}}(y') \right| = \frac{\varphi_{\mathrm{D}} - \varphi_0}{q R^*} \frac{\sinh\left[ (L_{\mathrm{D}} - y') / l^* \right]}{\cosh\left( L_{\mathrm{D}} / l^* \right)}, \tag{S11}$$

$$\left| I_{\mathrm{t}}(y') \right| = \frac{\varphi_{\mathrm{D}} - \varphi_0}{q R_{\mathrm{Drain}} L_{\mathrm{D}}} \frac{\cosh\left[ (L_{\mathrm{D}} - y') / l^* \right]}{\cosh\left( L_{\mathrm{D}} / l^* \right)}, \tag{S12}$$

$$R^* = \sqrt{R_{\mathrm{Drain}} R_{\mathrm{N}}} , \tag{S13}$$

$$l^* = L_{\mathrm{D}} \sqrt{\frac{R_{\mathrm{Drain}}}{R_{\mathrm{N}}}} , \tag{S14}$$

where $\varphi_0$ is the potential of the left-hand edge of the $n^+$-Si region ($y' = 0$), $\varphi_{\mathrm{D}}$ is the potential of the FM electrode above the MgO tunnel barrier, $R^*$ is the effective resistance of the D electrode, and $l^*$ expresses the length scale of the current crowding. Using Eqs. (S10) – (S14), $V_{DS}$ and $V_D$ are given by:

$$\begin{aligned} V_{DS} &= I_{DS}\left(R_{Source} + R_{ch}\right) + \left(\varphi_D - \varphi_0\right)/q \\ &= I_{DS}\left(R_{Source} + R_{ch} + R^* \frac{\cosh(L_D / l^*)}{\sinh(L_D / l^*)}\right), \end{aligned} \tag{S15}$$

$$\begin{aligned} V_D &= \left(\varphi_D - \varphi(L_D)\right)/q \\ &= I_{DS} R^* \frac{1}{\sinh(L_D / l^*)}, \end{aligned} \tag{S16}$$

Thus, $\Delta V_{DS}$ and $\Delta V_D$ are given by;

$$\Delta V_{DS} = I_{DS} \frac{\cosh(L_D / l^*)}{\sinh(L_D / l^*)} \Delta R^*, \tag{S17}$$

$$\Delta V_D = I_{DS} \frac{1}{\sinh(L_D / l^*)} \Delta R^*, \tag{S18}$$

where $\Delta R^*$ is the change in resistance due to the spin-valve effect defined by $R^* = R_0^* + \Delta R^*$ and $R_0^*$ is the effective resistance of the D junction in the parallel magnetization configuration. Equations (S17) and (S18) indicate that the amplitude of the spin-valve signal in $V_D$ is always smaller than that in $V_{DS}$, and the signal ratio $\Delta V_D / \Delta V_{DS}$ is given by;

$$\begin{aligned} \frac{\Delta V_D}{\Delta V_{DS}} &= \frac{1}{\cosh(L_D / l^*)} \\ &= \frac{1}{\cosh(\sqrt{R_N / R_{Drain}})}. \end{aligned} \tag{S19}$$

Equation (S19) means that the ratio $R_N / R_{Drain}$ is the characteristic value that determines the current crowding [S5]. Figure S2(a) shows the signals ratio $\Delta V_D / \Delta V_{DS}$ calculated by Eq. (S19) as a function of $R_N / R_{Drain}$. As the ratio $R_N / R_{Drain}$ increases, the signal ratio $\Delta V_D / \Delta V_{DS}$ decreases. When $R_N << R_{Drain}$, $\Delta V_D / \Delta V_{DS} \sim 1$ that is the same in the basic theory. In our spin MOSFETs, $R_N$ =

19.8 Ω, using $\rho_n$ = 0.89 mΩcm, $t_n$ = 5 nm, $W_{\mathrm{ch}}$ = 180 μm, and $L_{\mathrm{D}}$ = 2 μm. On the other hand, $R_{\mathrm{Drain}} = 11$–17 Ω was estimated from the tunnel diode devices having the same tunnel structure as that in the spin MOSFET [S6]. Since $R_{\mathrm{N}} / R_{\mathrm{Drain}}$ = 1.2–1.8, the signal ratio $\Delta V_{\mathrm{D}} / \Delta V_{\mathrm{DS}}$ = 0.5–0.6 is expected, which agrees with our experimental result $\Delta V_{\mathrm{D}} / \Delta V_{\mathrm{DS}}$ ~ 0.5. To show the origin of the signal difference clearly, we calculate a tunnel current distribution $I_{\mathrm{t}}(y')/I_{\mathrm{DS}}$ and a potential distribution $\varphi(y')$ in the $n^+$-Si region in Figs. S2(b) and (c), respectively, using $R_{\mathrm{Drain}}$= 14 Ω, $R_{\mathrm{N}}$ = 20 Ω, and $L_{\mathrm{D}}$ = 2 μm. In this situation, the current crowding occurs at $y'=0$ as shown in Fig. S2(b) and the potential $\varphi(L_{\mathrm{D}})$ at $y'=L_{\mathrm{D}}$ is in the middle of $\varphi_0$ and $\varphi_{\mathrm{D}}$ as shown in Fig. S2(c). Thus, $\Delta V_{\mathrm{DS}}$ shows the whole spin-valve signal produced by the tunnel junction, while $\Delta V_{\mathrm{D}}$ shows a half of the whole spin-valve signal. Therefore, we concluded that the signal amplitude difference between $\Delta V_{\mathrm{SD}}$ and $\Delta V_{\mathrm{D}}$ comes from the current crowding at the left-hand edge of the D electrode.

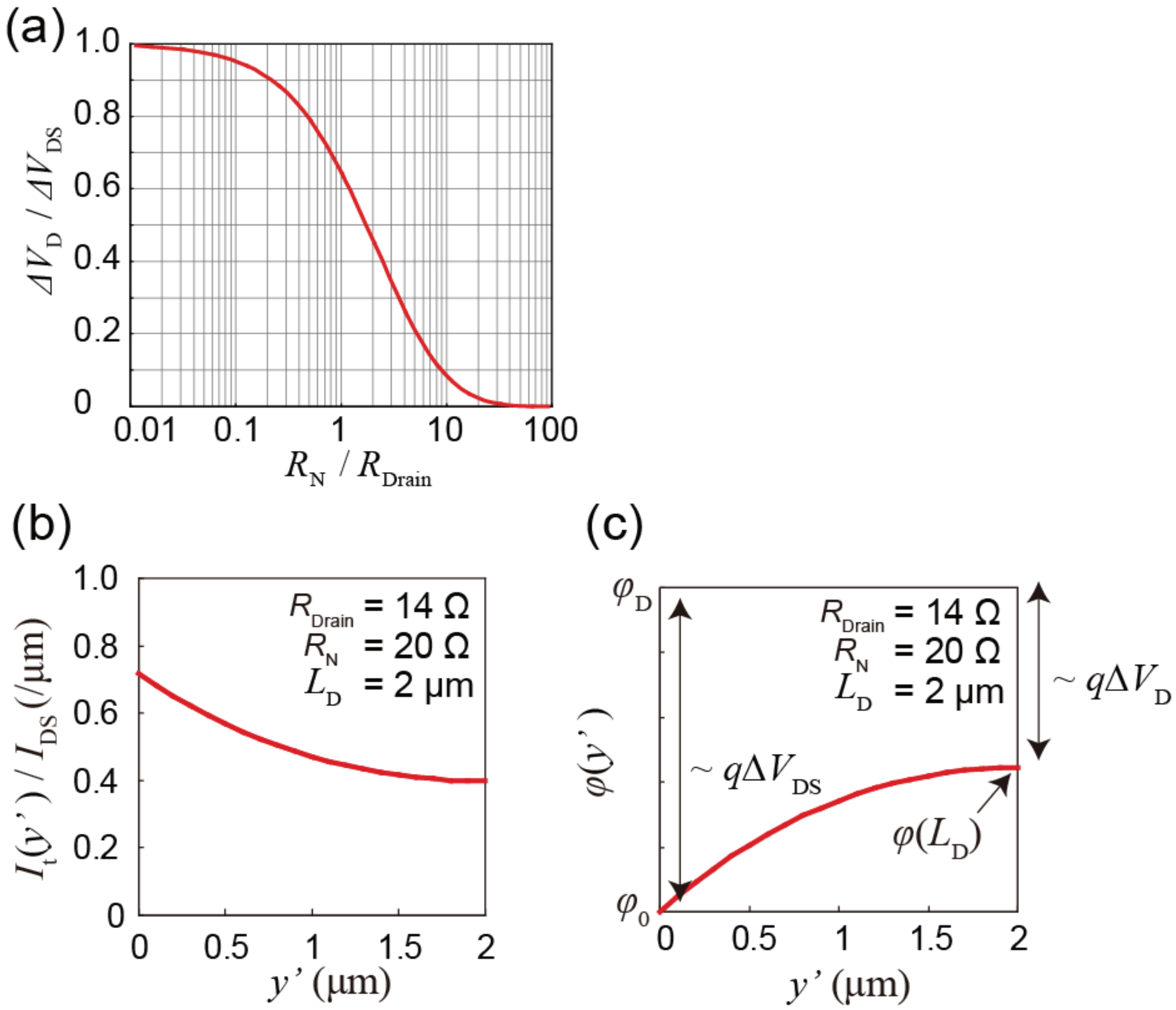

Figure S2 (a) Signal ratio $\mathit{\Delta} V_{\mathrm{D}} / \mathit{\Delta} V_{\mathrm{DS}}$ calculated using Eq. (S19) as a function of $R_{\mathrm{N}} / R_{\mathrm{Drain}}$. (b) Tunnel current distribution $I_{\mathrm{t}}(y') / I_{\mathrm{DS}}$ and (c) potential distribution $\varphi(y')$ calculated using Eq. (S11) and (S10), respectively, assuming $R_{\mathrm{Drain}}$= 14 Ω, $R_{\mathrm{N}}$ = 20 Ω, and $L_{\mathrm{D}}$ = 2 μm. In (c), double headed arrows indicate the voltage drop related to the signal $\mathit{\Delta} V_{\mathrm{DS}}$ and $\mathit{\Delta} V_{\mathrm{D}}$.

## S2. Derivation of the 2T Hanle expression

In this section, the 2T Hanle expression (Eq. (4a) in the main manuscript) is derived. The general solution of the spin drift-diffusion equation (Eq. (2) in the main manuscript) is as follows;

$$S(y)=\begin{cases} S_{\mathrm{A}}\exp\left(\dfrac{y+L_{\mathrm{S}}}{\lambda_{\mathrm{ch}}}\right) & (y\le -L_{\mathrm{S}}) \\ S_{\mathrm{B}}\exp\left(-\dfrac{y+L_{\mathrm{S}}}{\lambda_{\mathrm{n}}}\right)+S_{\mathrm{C}}\exp\left(\dfrac{y}{\lambda_{\mathrm{n}}}\right) & (-L_{\mathrm{S}}\le y\le 0) \\ S_{\mathrm{D}}\exp\left(-\dfrac{y}{\lambda_{\mathrm{ch}}^{\mathrm{d}}}\right)+S_{\mathrm{E}}\exp\left(\dfrac{y-L_{\mathrm{ch}}}{\lambda_{\mathrm{ch}}^{\mathrm{u}}}\right) & (0\le y\le L_{\mathrm{ch}}) \\ S_{\mathrm{F}}\exp\left(-\dfrac{y-L_{\mathrm{ch}}}{\lambda_{\mathrm{n}}}\right)+S_{\mathrm{G}}\exp\left(\dfrac{y-L_{\mathrm{ch}}-L_{\mathrm{D}}}{\lambda_{\mathrm{n}}}\right) & (L_{\mathrm{ch}}\le y\le L_{\mathrm{ch}}+L_{\mathrm{D}}) \\ S_{\mathrm{H}}\exp\left(-\dfrac{y-L_{\mathrm{ch}}-L_{\mathrm{D}}}{\lambda_{\mathrm{ch}}}\right) & (L_{\mathrm{ch}}+L_{\mathrm{D}}\le y) \end{cases} \quad \text{(S20)}$$

, where $S_{\mathrm{A}}$ – $S_{\mathrm{H}}$ [/cm$^2$] are the constants expressing the sheet spin density, $\lambda_{\mathrm{n}}=\sqrt{D_e^{\mathrm{n}}\tau_{\mathrm{S}}^{\mathrm{ch}}}$ and $\lambda_{\mathrm{ch}}=\sqrt{D_e^{\mathrm{ch}}\tau_{\mathrm{S}}^{\mathrm{ch}}}$ are the spin diffusion length in the $n^+$-Si region and the 2D channel region, respectively, and $\lambda_{\mathrm{ch}}^{\mathrm{u}}$ and $\lambda_{\mathrm{ch}}^{\mathrm{d}}$ are the up-stream and down-stream spin drift diffusion lengths, respectively [S7], defined by Eq. (4d) in the main manuscript. The constants $S_{\mathrm{A}}$ – $S_{\mathrm{H}}$ are determined by the boundary conditions at $y = -L_{\mathrm{S}}$, 0, $L_{\mathrm{ch}}$, and $L_{\mathrm{ch}} + L_{\mathrm{D}}$, where continuity is required for the spin current but not for the spin density $S$ due to the different carrier densities between the $n^+$-Si region and the 2D channel region. Thus, at the boundaries, we introduce the continuity of the spin accumulation $\Delta E=(E_+-E_-)/2$ that is the half of the difference in the electro-chemical potential between up-spin ($E_+$) and down-spin ($E_-$) electrons. The relation between $\Delta E_j$ and $S_j$ ($j$ = A – H) is expressed in each region as follows;

$$\begin{cases} S_{\mathrm{B}} = N_{\mathrm{n}} t_{\mathrm{n}} \Delta E_{\mathrm{B}} \\ S_{\mathrm{C}} = N_{\mathrm{n}} t_{\mathrm{n}} \Delta E_{\mathrm{C}} \\ S_{\mathrm{F}} = N_{\mathrm{n}} t_{\mathrm{n}} \Delta E_{\mathrm{F}} \\ S_{\mathrm{G}} = N_{\mathrm{n}} t_{\mathrm{n}} \Delta E_{\mathrm{G}} \end{cases}, \tag{S21}$$

in the $n^+$-Si region, and

$$\begin{cases} S_{\mathrm{A}} = N_{\mathrm{ch}} \Delta E_{\mathrm{A}} \\ S_{\mathrm{D}} = N_{\mathrm{ch}} \Delta E_{\mathrm{D}} \\ S_{\mathrm{E}} = N_{\mathrm{ch}} \Delta E_{\mathrm{E}} \\ S_{\mathrm{H}} = N_{\mathrm{ch}} \Delta E_{\mathrm{H}} \end{cases}, \tag{S22}$$

in the 2D channel region, where $\Delta E_{\mathrm{A}}$ – $\Delta E_{\mathrm{H}}$ are constants expressing the spin accumulation, $t_{\mathrm{n}}$ is the thickness of the $n^+$-Si region, and $N_{\mathrm{n}}$ [$\mathrm{cm}^{-3}\mathrm{J}^{-1}$] and $N_{\mathrm{ch}}$ [$\mathrm{cm}^{-2}\mathrm{J}^{-1}$] are the density of states (DOS) of the $n^+$-Si region and the 2D channel region, respectively. Consequently, the boundary conditions of the chemical potential are given by;

$$\Delta E_{\mathrm{A}} = \Delta E_{\mathrm{B}} + \varepsilon^{\mathrm{S}} \Delta E_{\mathrm{C}}, \tag{S23a}$$

$$\varepsilon^{\mathrm{S}} \Delta E_{\mathrm{B}} + \Delta E_{\mathrm{C}} = \Delta E_{\mathrm{D}} + \gamma^{\mathrm{u}} \Delta E_{\mathrm{E}}, \tag{S23b}$$

$$\gamma^{\mathrm{d}} \Delta E_{\mathrm{D}} + \Delta E_{\mathrm{E}} = \Delta E_{\mathrm{F}} + \varepsilon^{\mathrm{D}} \Delta E_{\mathrm{G}}, \tag{S23c}$$

$$\varepsilon^{\mathrm{D}} \Delta E_{\mathrm{F}} + \Delta E_{\mathrm{G}} = \Delta E_{\mathrm{H}}, \tag{S23d}$$

where the decay parameters $\varepsilon^{\mathrm{S}}$, $\varepsilon^{\mathrm{D}}$, $\gamma^{\mathrm{u}}$, and $\gamma^{\mathrm{d}}$ are defined by;

$$\varepsilon^{\mathrm{S}} = \exp\left(-\frac{L_{\mathrm{S}}}{\lambda_{\mathrm{n}}}\right), \quad \varepsilon^{\mathrm{D}} = \exp\left(-\frac{L_{\mathrm{D}}}{\lambda_{\mathrm{n}}}\right), \tag{S24a}$$

$$\gamma^{\mathrm{u}} = \exp\left(-\frac{L_{\mathrm{ch}}}{\lambda_{\mathrm{ch}}^{\mathrm{u}}}\right), \quad \gamma^{\mathrm{d}} = \exp\left(-\frac{L_{\mathrm{ch}}}{\lambda_{\mathrm{ch}}^{\mathrm{d}}}\right). \tag{S24b}$$

Since the spin current is calculated by $qW_{\mathrm{ch}}\left[-D_e \partial/\partial y + v_{\mathrm{d}}\right]S(y)$, the boundary conditions of the spin current are given by;

$$-\frac{\Delta E_{\mathrm{A}}}{qr_{\mathrm{ch}}}=\frac{\Delta E_{\mathrm{B}}}{qr_{\mathrm{n}}}-\varepsilon^{\mathrm{S}}\frac{\Delta E_{\mathrm{C}}}{qr_{\mathrm{n}}}, \tag{S25a}$$

$$P_{\mathrm{S}}I_{\mathrm{DS}}=-\varepsilon^{\mathrm{S}}\frac{\Delta E_{\mathrm{B}}}{qr_{\mathrm{n}}}+\frac{\Delta E_{\mathrm{C}}}{qr_{\mathrm{n}}}+\frac{\Delta E_{\mathrm{D}}}{qr_{\mathrm{ch}}^{\mathrm{d}}}-\gamma^{\mathrm{u}}\frac{\Delta E_{\mathrm{E}}}{qr_{\mathrm{ch}}^{\mathrm{u}}}, \tag{S25b}$$

$$-\sigma^{\mathrm{P/AP}}P_{\mathrm{S}}I_{\mathrm{DS}}=-\gamma^{\mathrm{d}}\frac{\Delta E_{\mathrm{D}}}{qr_{\mathrm{ch}}^{\mathrm{d}}}+\frac{\Delta E_{\mathrm{E}}}{qr_{\mathrm{ch}}^{\mathrm{u}}}+\frac{\Delta E_{\mathrm{F}}}{qr_{\mathrm{n}}}-\varepsilon^{\mathrm{D}}\frac{\Delta E_{\mathrm{G}}}{qr_{\mathrm{n}}}, \tag{S25c}$$

$$\varepsilon^{\mathrm{D}}\frac{\Delta E_{\mathrm{F}}}{qr_{\mathrm{n}}}-\frac{\Delta E_{\mathrm{G}}}{qr_{\mathrm{n}}}=\frac{\Delta E_{\mathrm{H}}}{qr_{\mathrm{ch}}}, \tag{S25d}$$

where $q$ is the elementary charge, $\sigma^{\mathrm{P/AP}}=\pm 1$ is the sign parameter expressing the parallel ($\sigma^{\mathrm{P}}=+1$) and an antiparallel ($\sigma^{\mathrm{AP}}=-1$) magnetization configurations of the S and D electrodes, $r_{\mathrm{n}}=\rho_{\mathrm{n}}\lambda_{\mathrm{n}}/t_{\mathrm{n}}W_{\mathrm{ch}}$ and $r_{\mathrm{ch}}=R_{\mathrm{S}}\lambda_{\mathrm{ch}}/W_{\mathrm{ch}}$ are the spin resistance of the $n^{+}$-Si region and the 2D channel region, respectively, and $r_{\mathrm{ch}}^{\mathrm{d}}=R_{\mathrm{S}}\lambda_{\mathrm{ch}}^{\mathrm{u}}/W_{\mathrm{ch}}$ and $r_{\mathrm{ch}}^{\mathrm{u}}=R_{\mathrm{S}}\lambda_{\mathrm{ch}}^{\mathrm{d}}/W_{\mathrm{ch}}$ (Eqs. (5a) and (5b) in the main manuscript) are the up-stream and down-stream spin resistances in the 2D channel region, respectively. The left-hand side of Eqs. (S25b) and (S25c) are spin injection from the S electrode and spin extraction from the D electrode, respectively. In the above-mentioned derivations, the relations $\rho_{\mathrm{n}}=1/qD_{e}^{\mathrm{n}}N_{\mathrm{n}}$ and $R_{\mathrm{S}}=1/qD_{e}^{\mathrm{ch}}N_{\mathrm{ch}}$ were used. We should emphasize here that the up-stream and down-stream spin resistances, $r_{\mathrm{ch}}^{\mathrm{u}}$ and $r_{\mathrm{ch}}^{\mathrm{d}}$, are defined by the spin drift-diffusion lengths along the opposite directions, $\lambda_{\mathrm{ch}}^{\mathrm{d}}$ and $\lambda_{\mathrm{ch}}^{\mathrm{u}}$, respectively, (as shown in Eqs. (5a) and (5b) in the main manuscript). This is because the spin resistance is inversely proportional to its effective spin diffusion length when $\tau_{\mathrm{S}}$ is unchanged (see section S3 for the detail), and up- and down-stream spin drift-diffusion lengths are inversely proportional to each other ($\lambda_{\mathrm{ch}}^{\mathrm{u}}\lambda_{\mathrm{ch}}^{\mathrm{d}}=(\lambda_{\mathrm{ch}})^{2}$). Therefore, as the spin drift increases, the effective spin diffusion length increases, but the effective spin resistance decreases. Solving Eqs. (S23) – (S25) leads to the distribution of the spin accumulation voltages in Figs. 7(a) – 7(d) in the main manuscript.

To obtain the analytical expression of the 2T Hanle signals, it is helpful to define the spin accumulation $\Delta E_0$ at $y$ = 0 ($\Delta E_0$ = Eq. (S23b)) and $\Delta E_{\mathrm{L}}$ at $y = L_{\mathrm{ch}}$ ($\Delta E_{\mathrm{L}}$ = Eq. (S23c)). Using them, Eqs. (S23a) – (S23d) and (S25a) – (S25d) are simplified to the following four equations;

$$\Delta E_0 = \Delta E_{\mathrm{D}} + \gamma^{\mathrm{u}} \Delta E_{\mathrm{E}}, \tag{S26a}$$

$$\gamma^{\mathrm{d}} \Delta E_{\mathrm{D}} + \Delta E_{\mathrm{E}} = \Delta E_{\mathrm{L}}, \tag{S26b}$$

$$P_{\mathrm{S}} I_{\mathrm{DS}} = \frac{\Delta E_0}{q r_{\mathrm{NL}}^{(\mathrm{S})}} + \frac{\Delta E_{\mathrm{D}}}{q r_{\mathrm{ch}}^{\mathrm{d}}} - \gamma^{\mathrm{u}} \frac{\Delta E_{\mathrm{E}}}{q r_{\mathrm{ch}}^{\mathrm{u}}}, \tag{S26c}$$

$$-\sigma^{\mathrm{P/AP}} P_{\mathrm{S}} I_{\mathrm{DS}} = \frac{\Delta E_{\mathrm{L}}}{q r_{\mathrm{NL}}^{(\mathrm{D})}} - \gamma^{\mathrm{d}} \frac{\Delta E_{\mathrm{D}}}{q r_{\mathrm{ch}}^{\mathrm{d}}} + \frac{\Delta E_{\mathrm{E}}}{q r_{\mathrm{ch}}^{\mathrm{u}}}, \tag{S26d}$$

where $r_{\mathrm{NL}}^{(\mathrm{S})}$ and $r_{\mathrm{NL}}^{(\mathrm{D})}$ are the effective spin resistances of the left-hand side of the spin injection point ($y \le 0$) and the right-hand side of the spin detection point ($L_{\mathrm{ch}} \le y$), respectively, as defined in Eqs. (5c) and (5d) in the main manuscript. Equations (S26a) and (S26b) are the continuities of the chemical potential at $y$ = 0 and $y = L_{\mathrm{ch}}$, respectively, whereas Eqs. (S26c) and (S26d) are the continuities of the spin current at $y$ = 0 and $y = L_{\mathrm{ch}}$, respectively. By solving them, we obtain;

$$\frac{\Delta E_0}{q} = \frac{P_{\mathrm{S}} I_{\mathrm{SD}}}{X}\left[\left(\frac{1}{r_{\mathrm{NL}}^{(\mathrm{D})}} + \frac{1}{r_{\mathrm{ch}}^{\mathrm{u}}}\right) - \left(\frac{1}{r_{\mathrm{NL}}^{(\mathrm{D})}} - \frac{1}{r_{\mathrm{ch}}^{\mathrm{d}}}\right)\gamma^{\mathrm{u}}\gamma^{\mathrm{d}} - \sigma^{\mathrm{P/AP}}\left(\frac{1}{r_{\mathrm{ch}}^{\mathrm{u}}} + \frac{1}{r_{\mathrm{ch}}^{\mathrm{d}}}\right)\gamma^{\mathrm{u}}\right], \tag{S27a}$$

$$\frac{\Delta E_{\mathrm{L}}}{q} = \frac{P_{\mathrm{S}} I_{\mathrm{SD}}}{X}\left[\left(\frac{1}{r_{\mathrm{ch}}^{\mathrm{u}}} + \frac{1}{r_{\mathrm{ch}}^{\mathrm{d}}}\right)\gamma^{\mathrm{d}} - \sigma^{\mathrm{P/AP}}\left(\frac{1}{r_{\mathrm{NL}}^{(\mathrm{S})}} + \frac{1}{r_{\mathrm{ch}}^{\mathrm{d}}}\right) + \sigma^{\mathrm{P/AP}}\left(\frac{1}{r_{\mathrm{NL}}^{(\mathrm{S})}} - \frac{1}{r_{\mathrm{ch}}^{\mathrm{u}}}\right)\gamma^{\mathrm{u}}\gamma^{\mathrm{d}}\right], \tag{S27b}$$

$$X = \left(\frac{1}{r_{\mathrm{NL}}^{(\mathrm{S})}} + \frac{1}{r_{\mathrm{ch}}^{\mathrm{d}}}\right)\left(\frac{1}{r_{\mathrm{NL}}^{(\mathrm{D})}} + \frac{1}{r_{\mathrm{ch}}^{\mathrm{u}}}\right) - \left(\frac{1}{r_{\mathrm{NL}}^{(\mathrm{S})}} - \frac{1}{r_{\mathrm{ch}}^{\mathrm{u}}}\right)\left(\frac{1}{r_{\mathrm{NL}}^{(D)}} - \frac{1}{r_{\mathrm{ch}}^{\mathrm{d}}}\right)\gamma^{\mathrm{u}}\gamma^{\mathrm{d}}. \tag{S27c}$$

Consequently, the spin signals $\Delta V_{\mathrm{S}}^{\mathrm{P/AP}}$ and $\Delta V_{\mathrm{D}}^{\mathrm{P/AP}}$ observed by the S and D electrodes, respectively, are;

$$\begin{aligned}\Delta V_{\mathrm{S}}^{\mathrm{P/AP}} &= \mathrm{Re}\left[P_{\mathrm{S}}\Delta E_0 / q\right] \\ &= \mathrm{Re}\left[\frac{{P_{\mathrm{S}}}^2 I_{\mathrm{DS}}}{X}\left\{\left(\frac{1}{r_{\mathrm{NL}}^{(\mathrm{D})}}+\frac{1}{r_{\mathrm{ch}}^{\mathrm{u}}}\right)-\left(\frac{1}{r_{\mathrm{NL}}^{(D)}}-\frac{1}{r_{\mathrm{ch}}^{\mathrm{d}}}\right)\gamma^{\mathrm{u}}\gamma^{\mathrm{d}}-\sigma^{\mathrm{P/AP}}\left(\frac{1}{r_{\mathrm{ch}}^{\mathrm{u}}}+\frac{1}{r_{\mathrm{ch}}^{\mathrm{d}}}\right)\gamma^{\mathrm{u}}\right\}\right],\end{aligned}$$

(S28a)

$$\begin{aligned}\Delta V_{\mathrm{D}}^{\mathrm{P/AP}} &= -\sigma^{\mathrm{P/AP}}\,\mathrm{Re}[P_{\mathrm{S}}\Delta E_{\mathrm{L}} / q] \\ &= \mathrm{Re}\left[\frac{{P_{\mathrm{S}}}^2 I_{\mathrm{DS}}}{X}\left\{\left(\frac{1}{r_{\mathrm{NL}}^{(\mathrm{S})}}+\frac{1}{r_{\mathrm{ch}}^{\mathrm{d}}}\right)-\left(\frac{1}{r_{\mathrm{NL}}^{(\mathrm{S})}}-\frac{1}{r_{\mathrm{ch}}^{\mathrm{u}}}\right)\gamma^{\mathrm{u}}\gamma^{\mathrm{d}}-\sigma^{\mathrm{P/AP}}\left(\frac{1}{r_{\mathrm{ch}}^{\mathrm{u}}}+\frac{1}{r_{\mathrm{ch}}^{\mathrm{d}}}\right)\gamma^{\mathrm{d}}\right\}\right].\end{aligned}$$

(S28b)

In our spin MOSFET structure, $\Delta V_{\mathrm{S}}^{\mathrm{P/AP}}$ is undetectable due to the strong electric field beneath the S electrode [S1, S5]. Therefore, the 2T Hanle expression is given by;

$$\begin{aligned}\Delta V_{\mathrm{D}}^{\mathrm{2TH(P/AP)}} &= \frac{V_{\mathrm{D}}^{\mathrm{P/AP}} - \Delta V_{\mathrm{D}}^{\mathrm{AP/P}}}{2} \\ &= -\sigma^{\mathrm{P/AP}}\,\mathrm{Re}\left[\frac{{P_{\mathrm{S}}}^2 I_{\mathrm{DS}}}{X}\left(\frac{1}{r_{\mathrm{ch}}^{\mathrm{u}}}+\frac{1}{r_{\mathrm{ch}}^{\mathrm{d}}}\right)\gamma^{\mathrm{d}}\right].\end{aligned} \qquad \text{(S29)}$$

This is Eq. (4a) in the main manuscript.

**S3. Reason why the up- and down-stream spin resistances are proportional to the opposite spin drift-diffusion lengths**

Spin resistance $r_S$ was firstly introduced by Valet and Fert [S8, S9], expressing the ratio between the spin current $J_S$ [A/m$^2$] and the spin accumulation potential difference $\Delta E(0)$ induced at the surface ($x = 0$) of a non-magnetic material (NM);

$$\frac{\Delta E(0)}{q} = J_S r_S. \tag{S30}$$

Although $r_S$ has the unit of [Ωm$^2$], its physical meaning is completely different from that of the ordinary electrical resistance, which is the ratio between the charge current and the voltage drop along the current direction. In an all-metallic system, $r_S$ is expressed as follows [S9];

$$r_S = \rho_N \lambda_S, \tag{S31}$$

where $\rho_N$ is the electrical resistivity and $\lambda_S$ is the spin diffusion length of NM. Eq. (S31) means that $r_S$ is directly proportion to $\lambda_S$ when $\rho_N$ is unchanged.

Let us think about the situation that NM is located in $x \geq 0$ and the spin current $J_S$ is injected into NM at $x = 0$ as shown in Fig. S3(a). Injected spins are accumulated near the interface and such a non-equilibrium state spreads in the range of effective spin diffusion length $\lambda_S^{\mathrm{eff}}$. Since the injected spins are consumed only by the spin relaxation, we write the spin relaxation current $J_{\mathrm{relax}}(x)$ at $x$ as follows;

$$J_{\mathrm{relax}}(x) = \frac{qS(x)}{\tau_S}, \tag{S32}$$

where $S(x)$ [m$^{-3}$] is the spin density at $x$ and $\tau_S$ is the spin lifetime. From the spin current continuity, the injected spin $J_S$ and the total spin consumption should be equal;

$$J_S = \int_0^\infty J_{\text{relax}}(x)dx$$
$$= \frac{q}{\tau_S}\int_0^\infty S(x)dx \quad , \qquad \text{(S33)}$$
$$= \frac{qN}{\tau_S}\int_0^\infty \Delta E(x)dx$$

where $\Delta E(x)$ is the spin accumuration at $x$, $N$ [$\mathrm{m^{-3}J^{-1}}$] is the density of states in NM and the relation $S(x)/N = \Delta E(x)$ was used. The third row of Eq. (S33) means that the gray-colored area in Fig. S3(a) is independent of $\lambda_S^{\text{eff}}$. Thus, as $\lambda_S^{\text{eff}}$ becomes longer, $\Delta E(0)$ becomes smaller as shown in Fig. S3(b). In contrast, as $\lambda_S^{\text{eff}}$ becomes shorter, $\Delta E(0)$ becomes larger as shown in Fig. S3(c). This means that the total number of spins accumulated in NM depends only on $\tau_S$ and $J_S$ as shown in the second row in Eq. (S33). When the spin distribution $S(x)$ has the following form;

$$S(x) = S_0 \exp\left(-\frac{x}{\lambda_S^{\text{eff}}}\right), \qquad \text{(S34)}$$

where $S_0$ is spin density at the interface ($x$ = 0), Eq. (S33) is written by;

$$J_S = \frac{qN\lambda_S^{\text{eff}}}{\tau_S}\Delta E(0). \qquad \text{(S35)}$$

From the definition Eq. (S30), the spin resistance $r_S$ is given by

$$r_S = \frac{\tau_S}{q^2 N\lambda_S^{\text{eff}}}. \qquad \text{(S36)}$$

Thus, $r_S$ is *inversely* proportional to $\lambda_S^{\text{eff}}$ if $\tau_S$ is unchanged. This relation is seemingly contradictory to the well-known expression Eq. (S31), but it is not, as explained below. Using the Einstein's relation $\rho_N = 1/q^2 N D_e$, where $D_e$ is the diffusion coefficient of the electrons in NM, Eq. (S36) is rewritten by;

$$r_S = \rho_N \frac{D_e \tau_S}{\lambda_S^{\text{eff}}} = \rho_N \frac{{\lambda_S}^2}{\lambda_S^{\text{eff}}}, \qquad \text{(S37)}$$

where $\lambda_S = \sqrt{D_e \tau_S}$ was used. When there is no spin drift ($\lambda_S^{eff} = \lambda_S$), Eq. (S37) is identical to Eq. (S31). On the other hand, when there is spin drift ($\lambda_S^{eff} = \lambda_S^{u(d)}$), using the relation $\lambda_S^u \lambda_S^d = \lambda_S{}^2$, the up(down)-stream spin resistance $r_S^{u(d)}$ is obtained;

$$r_S^{u(d)} = \rho_N \frac{\lambda_S{}^2}{\lambda_S^{u(d)}} = \rho_N \lambda_S^{d(u)} . \qquad \text{(S38)}$$

Thus, up(down)-stream spin resistance is proportional to down(up)-stream spin drift-diffusion length. Equation (S38) is equivalent to Eqs. (5a) and (5b) in the main manuscript. Note that Eq. (S38) [Eqs. (5a) and (5b)] is the extension of the concept of spin resistance [S9] to the general situation where both diffusion and drift exist.

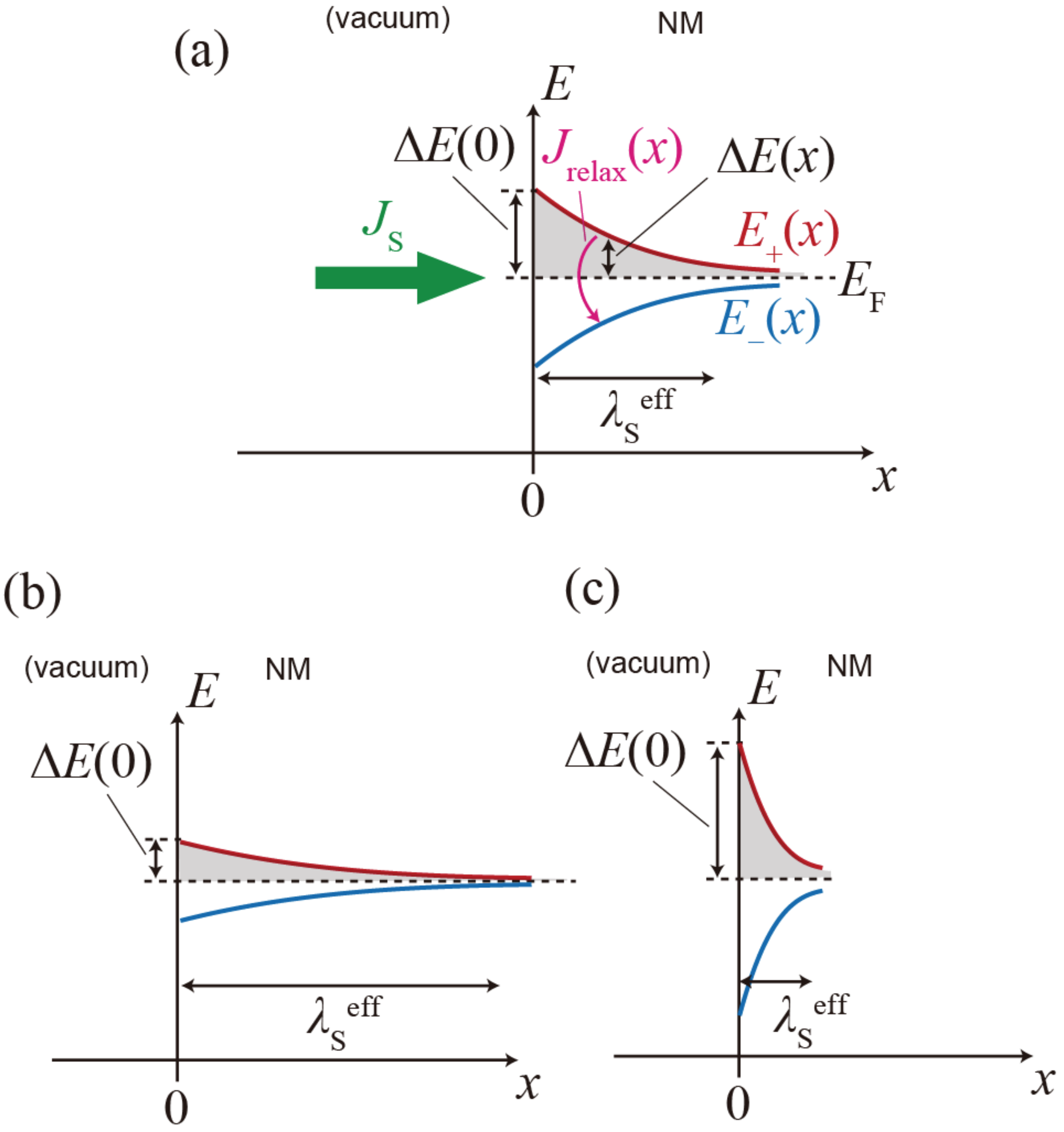

Fig. S3 (a) Potential distribution at NM surface with the spin injection $J_S$. $E_{+(-)}$ is the chemical potential of the up(down) spin electron, $\Delta E(0) = (E_+(0) - E_-(0))/2$, and $E_F$ is the Fermi level. $J_{relax}$ is the spin relaxation current and $\lambda_S^{eff}$ is the effective spin diffusion length. (b) As $\lambda_S^{eff}$ gets longer, $\Delta E(0)$ gets smaller and (c) as $\lambda_S^{eff}$ gets shorter, $\Delta E(0)$ gets larger, because the gray-colored area is independent of $\lambda_S^{eff}$ as shown in Eq. (S33).

**S4. Diffusion coefficient and momentum lifetime in the inversion channel**

To estimate the electron diffusion coefficient $D_e^{\mathrm{ch}}$ and electron momentum lifetime $\tau^{\mathrm{ch}}$ in the Si 2D inversion channel, we calculated the subband structure in the 2D inversion channel by solving Poisson's and Schrödinger's equations self-consistently in the same way as described in our previous paper [S1]. Since the electron mobility is smaller than that in our previous paper, especially in the low field range, the Coulomb scattering is taken into account in addition to the phonon and surface roughness scatterings.

First, the energy subband structure and carrier distribution in the 2D inversion channel were calculated for various gate electric fields $E_{\mathrm{ox}}$ = 0.1 – 5 MV/cm, assuming the thickness of the Si $t_{\mathrm{Si}}$ = 8 nm and the accepter doping concentration $N_{\mathrm{A}}$ = $1\times10^{15}$ cm$^{-3}$. Figure S4(a) shows the result at $E_{\mathrm{ox}}$ = 5 MV/cm, in which black and green solid curves are conduction band bottom $E_{\mathrm{C}}(z)$ and electron carrier density distribution $n(z)$, respectively, a black dashed line is the Fermi level $E_{\mathrm{F}}$, and red and blue lines are eigenenergies $E_i^{(v)}$ of the $i$th subband in the $v$-fold ($v$ = 2 or 4, respectively) valley. Since $t_{\mathrm{Si}}$ (= 8 nm) is thinner than that in our previous study (15 nm), the energy splitting between the higher energy subbands ($i$ > 2) is larger. Figure S4(b) shows $E_{\mathrm{F}}$, $E_i^{(2)}$, and $E_i^{(4)}$ plotted as a function of the effective field $E_{\mathrm{eff}} = q/\varepsilon_{\mathrm{Si}}\left(N_{\mathrm{S}}/2 - N_{\mathrm{D}}t_{\mathrm{Si}}\right)$ [S10], where black dashed, red solid, and blue solid curves are $E_{\mathrm{F}}$, $E_i^{(2)}$ ($i$ = 0 – 4), and $E_i^{(4)}$ ($i$ = 0 and 1), respectively. From the relation between $E_{\mathrm{F}}$ and $E_0^{(2)}$, the conduction band is degenerated at $V_{\mathrm{GS}}$ = 80 V, but not at $V_{\mathrm{GS}}$ = 40 V. Figure S4(c) shows the normalized electron population $N_i^{(v)}$ in each subband as a function of $E_{\mathrm{eff}}$, where red, pale red, and orange regions are $N_0^{(2)}$, $N_1^{(2)}$, and $\sum_{i=2-9} N_i^{(2)}$, respectively, and blue and dark blue regions are $N_0^{(4)}$ and $\sum_{i=1-9} N_i^{(4)}$, respectively. The electron population is changed greatly by $V_{GS}$; 55% of the total electron concentration is in the $E_0^{(2)}$ energy at $V_{GS}$ = 80V, whereas 40% of the total electron concentration is in each of the $E_0^{(2)}$ and $E_0^{(4)}$ energy at $V_{GS}$ = 40V.

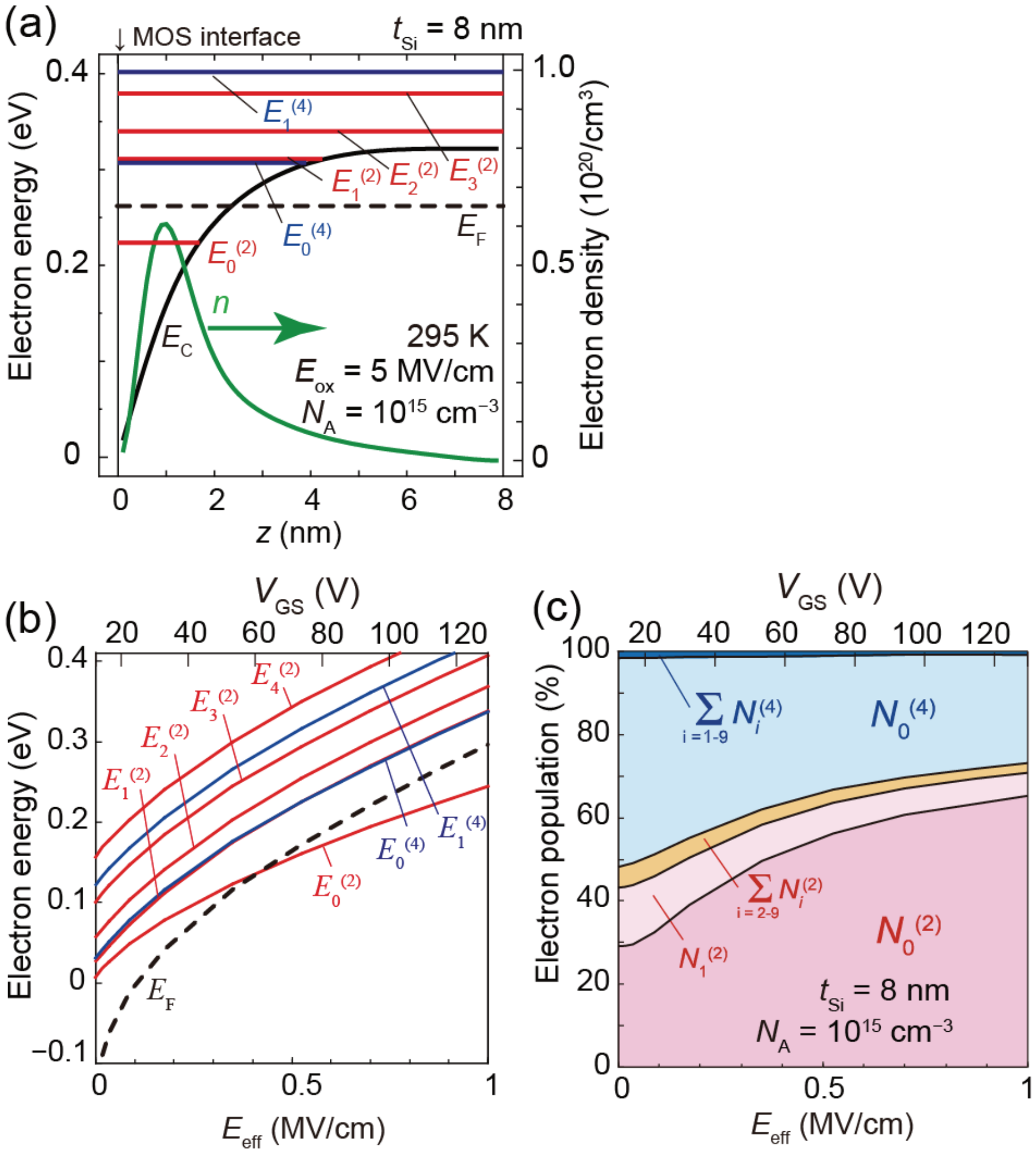

Figure S4 (a) Energy subband structure and carrier distribution in the inversion channel calculated at $E_{ox}$ = 5 MV/cm when the thickness of the Si $t_{Si}$ = 8 nm and the accepter doping concentration $N_A$ = $1 \times 10^{15}$ cm$^{-3}$. Black and green solid curves are the conduction band bottom $E_C(z)$ and carrier density distribution $n(z)$, respectively, a black dashed line is the Fermi level $E_F$, and red and blue lines are eigenenergies $E_i^{(v)}$ of the $i$th subband in the $v$-fold ($v$ = 2 or 4, respectively) valley. (b) $E_F$, $E_i^{(2)}$, and $E_i^{(4)}$ as a function of $E_{eff}$, where black dashed, red solid, and blue solid curves are $E_F$, $E_i^{(2)}$ ($i$ = 0–4), and $E_i^{(4)}$ ($i$ = 0 and 1), respectively. (c) Normalized electron population in the subband structure as a function of the effective electric field $E_{eff}$, where red, pale red, and orange regions are $N_0^{(2)}$, $N_1^{(2)}$, and $\Sigma_{i=2-9} N_i^{(2)}$, respectively, and blue and dark blue regions are $N_0^{(4)}$ and $\Sigma_{i=1-9} N_i^{(4)}$, respectively.

From these calculation results, the effective mobility $\mu_{\mathrm{ph+sr}}$ determined by phonon and surface roughness scatterings is obtained by averaging the population-weighted sum of the electron mobility of each subband;

$$\mu_{\mathrm{ph+sr}} = \sum_i N_i^{(2)} \mu_i^{(2)} / N_{\mathrm{S}} + \sum_i N_i^{(4)} \mu_i^{(4)} / N_{\mathrm{S}} , \quad \text{(S39a)}$$

$$\mu_i^{(v)} = \frac{q\tau_i^{(v)}}{m_{\mathrm{c}}^{(v)}} , \quad \text{(S39b)}$$

where $N_{\mathrm{S}}$ is the total sheet carrier density, $\tau_i^{(v)}$ is the momentum lifetime in the $i$th subband of the $v$-valley determined by phonon and surface roughness scattering (see S.M. of ref. [S1]), $m_{\mathrm{c}}^{(2)} = 0.19\ m_0$, $m_{\mathrm{c}}^{(4)} = 0.315\ m_0$ , and $m_0$ (= $9.11\times10^{-31}$ kg) is the free electron mass in a vacuum. Then, the Coulomb scattering is taken into account by the following form [S11].

$$\frac{1}{\mu_{\mathrm{total}}} = \frac{1}{\mu_{\mathrm{Coulomb}}} + \frac{1}{\mu_{\mathrm{ph+sr}}} , \quad \text{(S40a)}$$

$$\mu_{\mathrm{Coulomb}} = \alpha N_{\mathrm{S}} , \quad \text{(S40b)}$$

where $\mu_{\mathrm{Coulomb}}$ is the mobility determined by the Coulomb scattering and $\alpha$ is a fitting parameter. Equation (S40b) expresses the screening of the Coulomb scattering by the accumulated electrons [S10]. The scattering parameters (see S.M. of ref. [S1]) $D_f$, $D_g$, $D_{\mathrm{ac}}$, $E_f$, $E_g$, $\Delta\Lambda$, and $\alpha$ were determined so that the experimental mobility values in Fig. 5(c) in the main manuscript are reproduced. Figure S5(a) shows the calculation results, in which the green open diamonds are the experimental mobility estimated from the Hall measurement, the green solid curve is $\mu_{\mathrm{ph+sr}}$ calculated using Eq. (S39a), the red solid curve is $\mu_{\mathrm{Coulomb}}$ calculated using Eq. (S40b), and the blue solid curve is $\mu_{\mathrm{total}}$ calculated using Eq. (S40a). The determined parameters are listed in Table. S1.

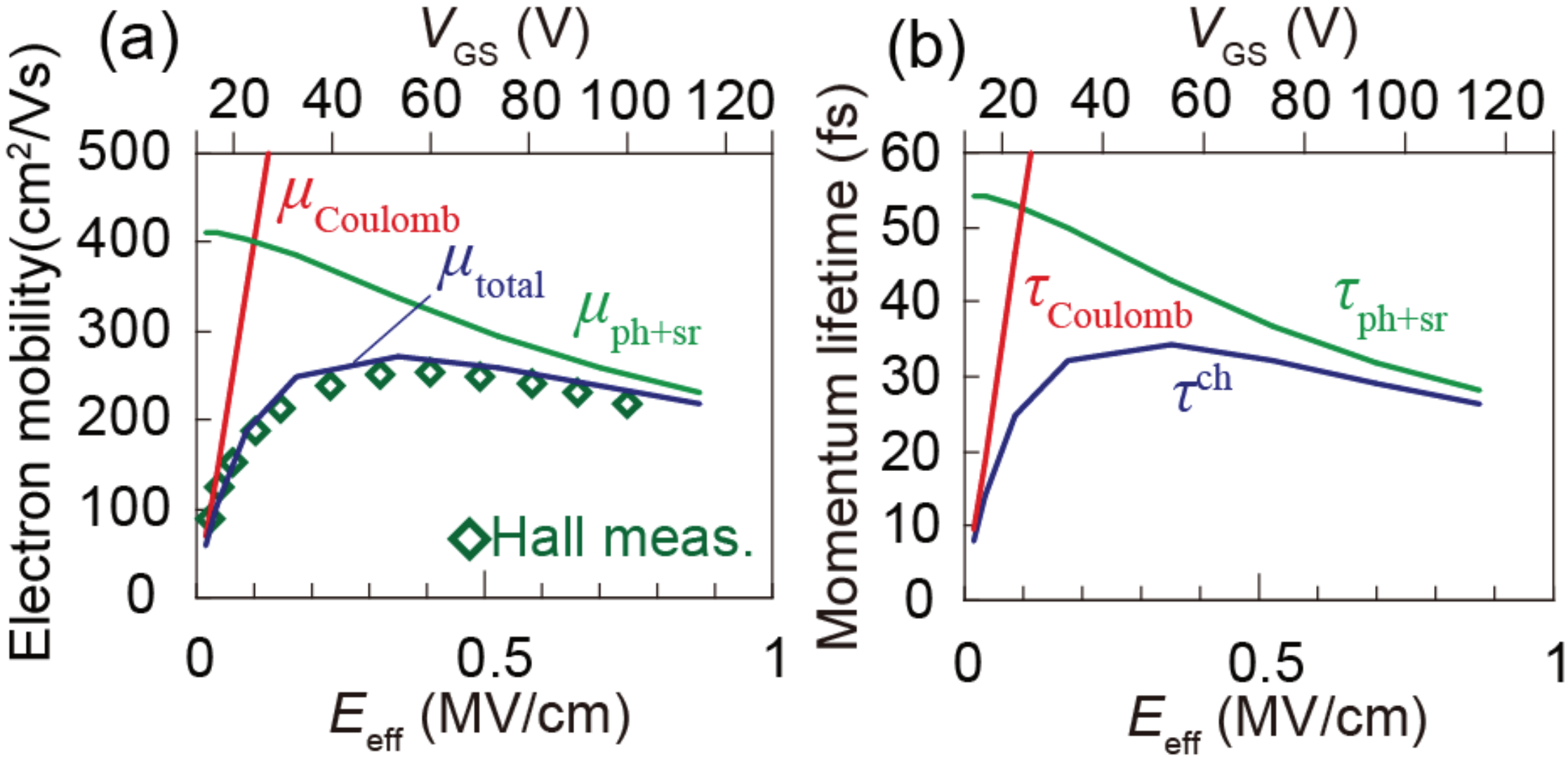

Figure S5 (a) Electron nobilities as functions of $E_{\text{eff}}$. Green diamonds are the experimental mobility estimated from the Hall measurement. The green, red, and blue solid curves are $\mu_{\text{ph+sr}}$, $\mu_{\text{Coulomb}}$, and $\mu_{\text{total}}$, calculated using Eqs. (S39a), (S40b), and (S40a), respectively. (b) Electron momentum lifetimes as functions of $E_{\text{eff}}$. The green, red, and blue solid lines are $\tau_{\text{ph+sr}}$, $\tau_{\text{Coulomb}}$, and $\tau^{\text{ch}}$ calculated using Eqs. (S41), (S42a), and (S43), respectively.

Table S1 Scattering parameters which were determined so that the experimental mobility in Fig. S5(a) is reproduced.

| Parameter | Value |
|---|---|
| $D_f$ | $15\times10^{8}$ eV/cm |
| $D_g$ | $15\times10^{8}$ eV/cm |
| $D_{\text{ac}}$ | 12 eV |
| $E_f$ | 59 eV |
| $E_g$ | 63 eV |
| $\Delta\Lambda$ | $25\times10^{-8}$ m$^2$ |
| $\alpha$ | $3\times10^{-10}$ cm$^4$/Vs |

Using these scattering parameters, the momentum lifetime $\tau_{\mathrm{ph+sr}}$ determined by the phonon and surface roughness scatterings is calculated by;

$$\tau_{\mathrm{ph+sr}} = \sum_i N_i^{(2)} \tau_i^{(2)} / N_{\mathrm{S}} + \sum_i N_i^{(4)} \tau_i^{(4)} / N_{\mathrm{S}} , \tag{S41}$$

and the momentum lifetime $\tau_{\mathrm{Coulomb}}$ determined by the Coulomb scattering is calculated by;

$$\tau_{\mathrm{Coulomb}} = \frac{m_{\mathrm{c}}^{*} m_0 \mu_{\mathrm{Coulomb}}}{q} , \tag{S42a}$$

$$m_{\mathrm{c}}^{*} = \left( \sum_i N_i^{(2)} / m_{\mathrm{c}}^{(2)} + \sum_i N_i^{(4)} / m_{\mathrm{c}}^{(4)} \right)^{-1} , \tag{S42b}$$

where $m_{\mathrm{c}}^{*}$ is the effective conductivity mass of electrons considering the population of the each subband. Finally, we get the effective momentum lifetime $\tau^{\mathrm{ch}}$ in the inversion channel;

$$\frac{1}{\tau^{\mathrm{ch}}} = \frac{1}{\tau_{\mathrm{Coulomb}}} + \frac{1}{\tau_{\mathrm{ph+sr}}} . \tag{S43}$$

Each momentum lifetime $\tau_{\mathrm{ph+sr}}$, $\tau_{\mathrm{Coulomb}}$, and $\tau^{\mathrm{ch}}$ are shown by green, red, and blue solid lines in Fig. S5(b), respectively. $\tau^{\mathrm{ch}}$ values are plotted in Fig. 5(e) in the main manuscript to compare the $\tau^{\mathrm{ch}}$ values with the spin lifetime in the channel $\tau_{\mathrm{S}}^{\mathrm{ch}}$.

On the other hand, the diffusion coefficient $D_{\mathrm{ph+sr}}$ determined by the phonon and surface roughness scatterings is calculated as follows [S1, S11];

$$D_{\mathrm{ph+sr}} = \sum_i N_i^{(2)} D_i^{(2)} / N_{\mathrm{S}} + \sum_i N_i^{(4)} D_i^{(4)} / N_{\mathrm{S}} , \tag{S44a}$$

$$D_i^{(v)} = \frac{kT\mu_i^{(v)}}{q} \left[ 1 + \exp(-\frac{E_{\mathrm{F}} - E_i^{(v)}}{qkT}) \right] \ln(1 + \exp(\frac{E_{\mathrm{F}} - E_i^{(v)}}{qkT})) . \tag{S44b}$$

Considering the fact that the Coulomb scattering is dominant at the lower gate electric field range, the diffusion coefficient $D_{\mathrm{Coulomb}}$ determined by the Coulomb scattering is calculated by the Einstein's relation for non-degenerated semiconductors [S11];

$$qD_{\text{Coulomb}} = kT\mu_{\text{Coulomb}} . \tag{S45}$$

Finally, we get the effective diffusion coefficient $D_e^{\text{ch}}$ in the inversion channel;

$$\frac{1}{D_e^{\text{ch}}} = \frac{1}{D_{\text{Coulomb}}} + \frac{1}{D_{\text{ph+sr}}} . \tag{S46}$$

$D_e^{\text{ch}}$ values calculated by Eq. (S46) are shown by the blue solid curve in Fig. 5(d) in the main manuscript.

## S5. Solution for a single channel material and uniform electric field

When the channel is made of a single material and the lateral electric field along the $y$ direction is uniform in the entire channel, $r_{\rm NL}^{\rm (S)}$ and $r_{\rm NL}^{(D)}$ of Eq. (S29) are replaced by $r_{\rm ch}^{\rm u}$ and $r_{\rm ch}^{\rm d}$, respectively. In this case, Eq. (S29) is identical with the conventional expression of 2T Hanle signals [S12-S14];

$$\Delta V_{\rm D}^{\rm 2TH} = P_{\rm S}^2 I_{\rm DS}\frac{R_{\rm S}}{W_{\rm ch}}{\rm Re}\left[\left(\frac{1}{\lambda_{\rm ch}^{\rm u}}+\frac{1}{\lambda_{\rm ch}^{\rm d}}\right)^{-1}\exp\left(-\frac{L_{\rm ch}}{\lambda_{\rm ch}^{\rm d}}\right)\right]$$

$$= \frac{\lambda^{\rm drift}}{2}\frac{1}{\sqrt[4]{1+(\gamma H_{\perp}\tau_{\rm S}^{\rm drift})^2}}\exp\left(\frac{L_{\rm ch}}{2\Lambda}-\frac{L_{\rm ch}}{\lambda^{\rm drift}}\sqrt{\frac{\sqrt{(\gamma H_{\perp}\tau_{\rm S}^{\rm drift})^2+1}}{2}}\right)$$

$$\times\cos\left(\frac{{\rm Tan}^{-1}(\gamma H_{\perp}\tau_{\rm S}^{\rm drift})}{2}+\frac{L_{\rm ch}}{\lambda^{\rm drift}}\sqrt{\frac{\sqrt{1+(\gamma H_{\perp}\tau_{\rm S}^{\rm drift})^2}-1}{2}}\right), \qquad \text{(S47a)}$$

$$\tau_{\rm S}^{\rm drift} = \left(\frac{1}{\tau_{\rm S}^{\rm ch}}+\frac{v_{\rm d}^{2}}{4D_e^{\rm ch}}\right)^{-1}, \qquad \text{(S47b)}$$

$$\lambda^{\rm drift} = \sqrt{D_e^{\rm ch}\tau_{\rm S}^{\rm drift}}, \qquad \text{(S47c)}$$

where $\tau_{\rm S}^{\rm drift}$ and $\lambda^{\rm drift}$ are the effective spin lifetime and spin diffusion length, respectively, when the spin drift is taken into account. Figures S6(a) and (b) show fitting results using Eq. (S29) (= Eq. (4a) in the main manuscript) and Eq. (S47a), respectively, where red and blue solid curves are the experimental 2T Hanle signals with $V_{\rm GS}$ = 40 V and $I_{\rm DS}$ = 10 mA, black dashed curves are the fitting results, and black dotted curves are the envelope of the fitting functions. The envelope function was calculated by taking the absolute value of the complex function instead of the real part, because the absolute value of the complex spin density expresses the spin density whereas the real part expresses the projection of the spin density onto the $x$-axis [S5]. Our expression (Eq. (4a) and Eq. (S29)) perfectly reproduces the experimental signals as shown in Fig. S6(a), but the conventional expression (Eq. (S47a)) does not, as shown in Fig. S6(b): Although the oscillating behavior is reproduced by both functions, the envelope of their signals is completely different. This difference reflects the spin dephasing of the accumulated spins at $y = L_{ch}$ induced by the perpendicular magnetic field $H_{\perp}$. To explain this qualitatively, schematic images of the spin transport under the same $H_{\perp}$ and different lateral electric field ($F$) distribution are shown in Fig. S7. When $F$ is uniform in the entire channel (Fig. S7(b)), transported spins are drifted away into the nonlocal region ($L_{ch} < y$), i.e., they are not accumulated at $y = L_{ch}$. As a result, the spin dephasing at $y = L_{ch}$ is not so strong and the envelope becomes broader, as shown in Fig. S6(b). On the other hand, when $F$ is present only in the 2D channel region between the S and D electrodes (Fig. S6(a)), transported spins are accumulated at $y = L_{ch}$ beneath the D electrode, which leads to such steep reduction of the spins due to the decoherence of the accumulated spins, as shown in Fig. S6(a). These results strongly indicate that it is important to take into account the electric field distribution for the precise analysis of spin drift signals.

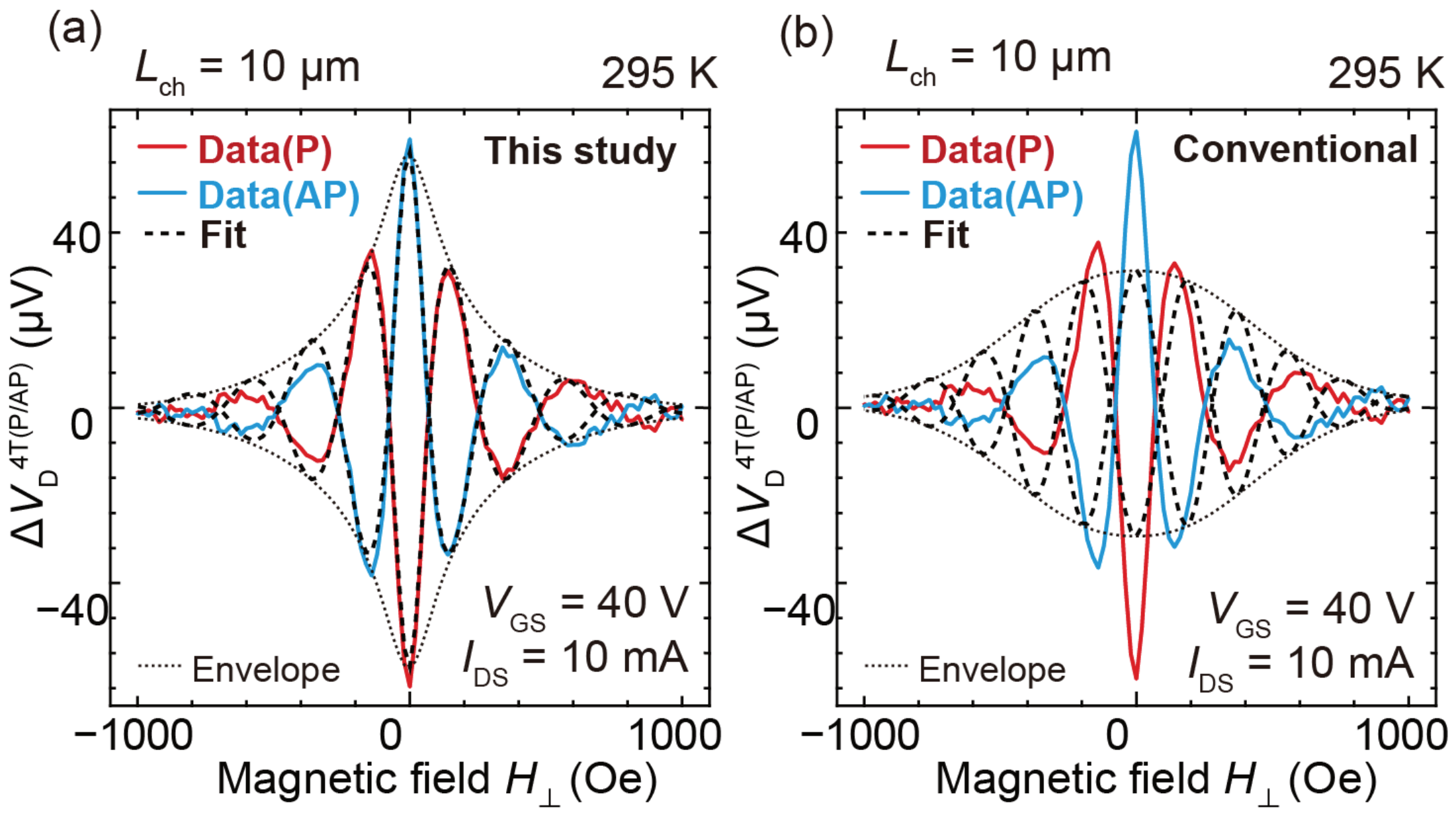

Figure S6 (a)(b) Fitting results of the Hanle signals using (a) Eq. (S29) (= Eq. (4a) in the main manuscript) and (b) Eq. (S47a). Red and blue solid curves are the experimental 2T Hanle signals with $V_{GS}$ = 40 V and $I_{DS}$ = 10 mA, black dashed curves are fitting results, and black dotted curves are the envelope of the fitting functions.

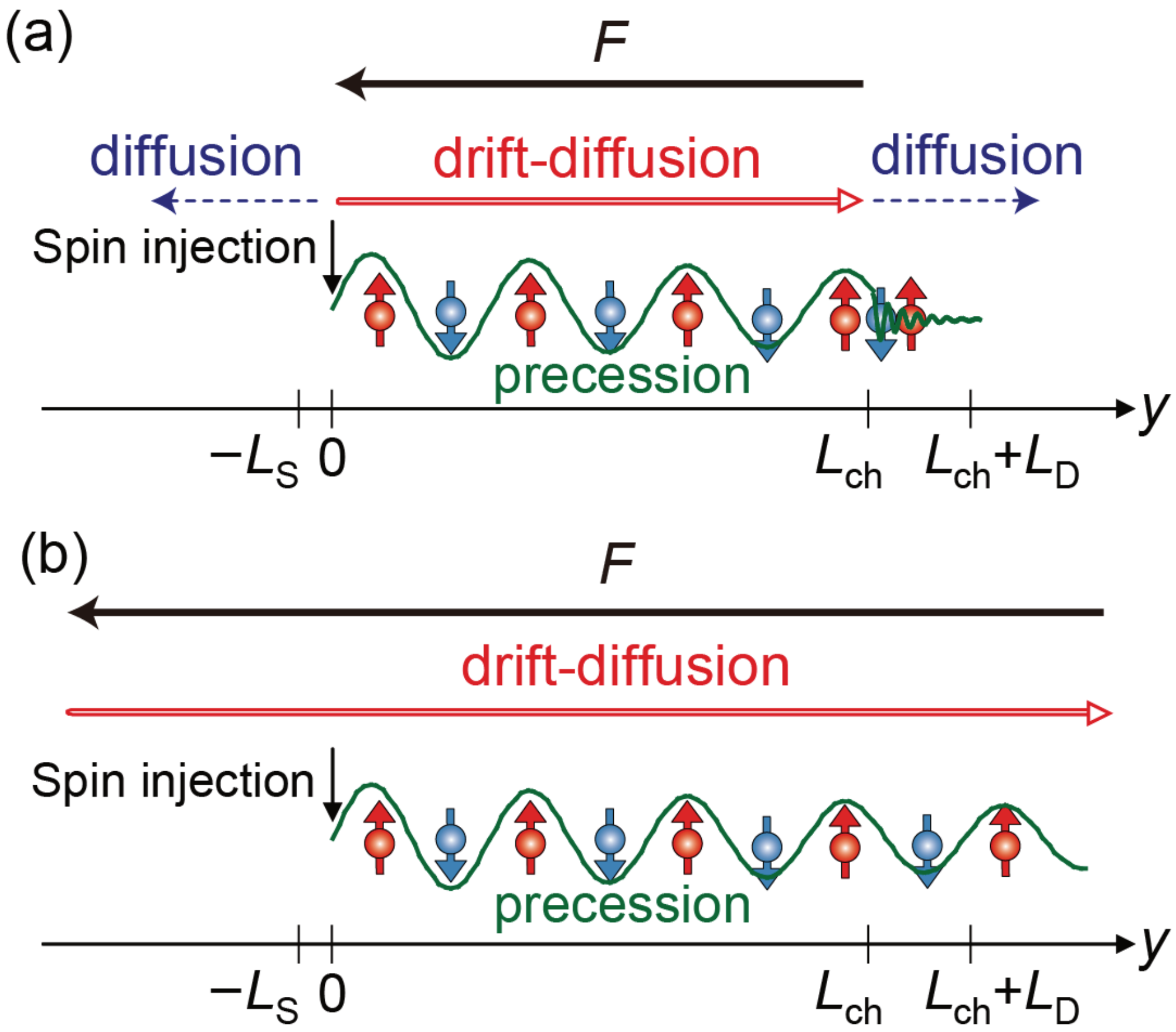

Figure S7 (a)(b) Schematic images of the spin transport under a perpendicular magnetic field $H_{\perp}$ with (a) a lateral electric field $F$ only in the 2D channel region between the S and D electrodes ($0 \leq y \leq L_{ch}$) and (b) a uniform electric field $F$ in the entire channel. Spin precession is symbolized by the up (red) and down (blue) spin electrons. Green solid lines express the spin density along the $x$ direction. Red lateral arrows express spin drift-diffusion assisted by $F$ and blue dashed arrows express spin diffusion. Transported spins are drifted away into the nonlocal region ($L_{ch} < y$) in (b), whereas they are accumulated at $y = L_{ch}$ in (a), which leads to the strong decoherence of the spins by $H_{\perp}$.

## S6. Introducing the conductivity matching condition into the 2T Hanle expression

In this section, we introduce the conductivity matching condition [S8, S9, S15, S16] into our 2T Hanle and spin-valve expressions (Eqs. (4a) and (6) in the main manuscript) to show why $r^{\mathrm{input}}$ and $r^{\mathrm{output}}$ determine the conductivity matching condition. Since the reduction of the spin injection efficiency due to the conductivity mismatch is cause by the back flow of the spin current from the $n^{+}$-Si under the S electrode to the S electrode, the spin current continuity at $y = 0$ (Eq. (S26c)) is replaced by

$$P_{\mathrm{S}} I_{\mathrm{DS}} - \frac{\Delta E_0}{q r_{\mathrm{B}}^{(\mathrm{S})}} = \frac{\Delta E_0}{q r_{\mathrm{NL}}^{(\mathrm{S})}} + \frac{\Delta E_{\mathrm{D}}}{q r_{\mathrm{ch}}^{\mathrm{d}}} - \gamma^{\mathrm{u}} \frac{\Delta E_{\mathrm{E}}}{q r_{\mathrm{ch}}^{\mathrm{u}}}, \tag{S48}$$

where $r_{\mathrm{B}}^{(\mathrm{S})}$ is the tunnel resistance of the S junction and the second term of the left-hand side expresses the back flow of the spin current from the $n^{+}$-Si to the S electrode. Similarly, since the reduction of the spin detection efficiency due to the conductivity mismatch is cause by the leakage of the spin current from the $n^{+}$-Si under the D electrode to the D electrode, the equation of the spin current continuity at $y = L_{\mathrm{ch}}$ (Eq. (S26d)) is replaced by

$$-\sigma^{\mathrm{P/AP}} P_{\mathrm{S}} I_{\mathrm{DS}} - \frac{\Delta E_L}{q r_{\mathrm{B}}^{(\mathrm{D})}} = \frac{\Delta E_{\mathrm{L}}}{q r_{\mathrm{NL}}^{(\mathrm{D})}} - \gamma^{\mathrm{d}} \frac{\Delta E_{\mathrm{D}}}{q r_{\mathrm{ch}}^{\mathrm{d}}} + \frac{\Delta E_{\mathrm{E}}}{q r_{\mathrm{ch}}^{\mathrm{u}}}, \tag{S49}$$

where $r_{\mathrm{B}}^{(\mathrm{D})}$ is the tunnel resistance of the D junction and the second term of the left-hand side expresses the leakage of the spin current from the $n^{+}$-Si under the D electrode to the D electrode. When $r_{\mathrm{B}}^{(\mathrm{D})}$ is high enough ($r_{\mathrm{B}}^{(\mathrm{D})} >> r^{\mathrm{output}}$), such spin current leakage can be neglected. By solving Eqs. (S26a), (S26b), Eq. (S48), and (S49), the following Eqs. (S50a), (S50b), and (50c) are alternative expressions of Eqs. (4a), (4b), and (6) in the main manuscript, respectively;

$$\varDelta V_{\mathrm{D}}^{\mathrm{4TH(P/AP)}} = -\sigma^{\mathrm{P/AP}} \,\mathrm{Re}\left[ \frac{P_{\mathrm{S}}^{\;2} I_{\mathrm{DS}}}{X'} \left( \frac{1}{r_{\mathrm{ch}}^{\mathrm{u}}} + \frac{1}{r_{\mathrm{ch}}^{\mathrm{d}}} \right) \gamma^{\mathrm{d}} \right], \tag{S50a}$$

$$X' = \left( \frac{1}{r_{\mathrm{NL}}^{(\mathrm{S})}} + \frac{1}{r_{\mathrm{ch}}^{\mathrm{d}}} + \frac{1}{r_{\mathrm{B}}^{(\mathrm{S})}} \right) \left( \frac{1}{r_{\mathrm{NL}}^{(\mathrm{D})}} + \frac{1}{r_{\mathrm{ch}}^{\mathrm{u}}} + \frac{1}{r_{\mathrm{B}}^{(\mathrm{D})}} \right) - \left( \frac{1}{r_{\mathrm{NL}}^{(\mathrm{S})}} - \frac{1}{r_{\mathrm{ch}}^{\mathrm{u}}} + \frac{1}{r_{\mathrm{B}}^{(\mathrm{D})}} \right) \left( \frac{1}{r_{\mathrm{NL}}^{(\mathrm{D})}} - \frac{1}{r_{\mathrm{ch}}^{\mathrm{d}}} + \frac{1}{r_{\mathrm{B}}^{(\mathrm{S})}} \right) \gamma^{\mathrm{u}} \gamma^{\mathrm{d}}, \tag{S50b}$$

$$\varDelta V^{\mathrm{2T}} = \left. \frac{2 P_{\mathrm{S}}^{\;2} I_{\mathrm{DS}}}{X'} \left( \frac{1}{r_{\mathrm{ch}}^{\mathrm{u}}} + \frac{1}{r_{\mathrm{ch}}^{\mathrm{d}}} \right) \gamma^{\mathrm{d}} \right|_{H_{\perp}=0} . \tag{S50c}$$

When the spin drift dominates the spin transport ($\lambda_{\mathrm{ch}}^{\mathrm{u}} << L_{\mathrm{ch}}$), Eq. (S50c) can be written by the following form;

$$\varDelta V^{\mathrm{2T}} \sim 2 \times P_{\mathrm{S}} \frac{r_{\mathrm{B}}^{(\mathrm{S})}}{r_{\mathrm{B}}^{(\mathrm{S})} + r^{\mathrm{input}}} I_{\mathrm{DS}} \times \frac{r^{\mathrm{input}}}{r_{\mathrm{ch}}^{\mathrm{d}}} \times \gamma^{\mathrm{d}} \times P_{\mathrm{S}} \frac{r_{\mathrm{B}}^{(\mathrm{D})}}{r_{\mathrm{B}}^{(\mathrm{D})} + r^{\mathrm{output}}} r^{\mathrm{output}} . \tag{S51}$$

Comparing Eq. (S51) with Eq. (7) in the main manuscript, one finds that the effective spin injection and detection polarization is $P_{\mathrm{S}}^{*} = P_{\mathrm{S}} r_{\mathrm{B}}^{(\mathrm{S})} / (r_{\mathrm{B}}^{(\mathrm{S})} + r^{\mathrm{input}})$ and $P_{\mathrm{S}}^{*} = P_{\mathrm{S}} r_{\mathrm{B}}^{(\mathrm{D})} / (r_{\mathrm{B}}^{(\mathrm{D})} + r^{\mathrm{output}})$, respectively. Therefore, $r_{\mathrm{B}}^{(\mathrm{S})} \sim r^{\mathrm{input}}$ and $r_{\mathrm{B}}^{(\mathrm{D})} \sim r^{\mathrm{output}}$ are the conductivity matching conditions, as mentioned in Section V in the main manuscript.